# Frequency-Space Mechanics: A Sequence and Coordinate-Free Representation for Protein Function Prediction

Author: Charles B Reilly

Affiliation: $CR_3$ - Reflexive Discovery, Cambridge, MA, USA

Correspondence: charles@reflexivediscovery.com

## Abstract

Protein function prediction is dominated by representations grounded in sequence and static structure, neither of which captures the collective vibrational dynamics through which proteins act. Here we introduce frequency-space mechanics, a representational framework in which a protein is encoded as a mechanical harmonics graph (MHG): nodes are vibrational modes derived from molecular dynamics, and edges are harmonic couplings weighted by octave alignment between mode frequencies. The representation is coordinate-free, sequence-independent, scale-invariant, and inhabits a latent mechanical space in which the original atomic coordinates have been projected out. The same construction applies to any system with a tractable eigendecomposition. Trained on 5,238 SwissProt proteins under a strict 30% sequence-identity split and using no sequence information, a graph neural network over static MHGs predicts GO molecular function terms across the ontology, demonstrating that vibrational physics alone encodes broad functional class. Kuramoto entrainment of the harmonic coupling graph, formally a Hamiltonian operation over mode frequencies and directly compatible with quantum annealing hardware, improves prediction for proteins whose function depends on collective conformational dynamics. On CLIC1, a fold- and function-switching chloride channel excluded from training, entrainment amplifies channel-activity signal 7.5-fold and antioxidant signal 2.4-fold, recovering both functional states from dynamics alone.

## Introduction

Protein function prediction has been transformed over the past decade by representations that capture sequence and structure with unprecedented fidelity. Language models trained on hundreds of millions of sequences now produce embeddings that recover remote homology, predict mutational effects, and seed generative design[1,2]. AlphaFold2 and its successors have effectively solved the static structure problem for the majority of natural proteins, providing high-confidence three-dimensional coordinates at proteome scale[3,4]. These advances have reshaped what is computationally tractable in molecular biology. Yet a representational gap remains. Sequence encodes the information available to evolution, and static structure encodes one snapshot of the conformational ensemble, but neither captures the collective vibrational dynamics through which proteins actually do their work. Catalysis, allostery, conformational selection, induced fit, and substrate channelling are all mechanical processes that unfold across the soft modes of a protein's vibrational spectrum. A representation that is blind to these dynamics is, by construction, blind to the physical mechanism of function.

Vibrational dynamics has been studied in proteins for decades. Normal mode analysis, elastic network models, Gaussian network models, and essential-dynamics covariance methods all extract mode spectra from structures or trajectories[5–9] and have been used productively to interpret allosteric communication, identify hinge regions, and predict conformational change. More recently, generative approaches have begun conditioning protein design directly on vibrational mode profiles, demonstrating that dynamics can serve as a tractable design target and that many distinct sequences can realise the same collective motion[10]. What has been missing is a representation of vibrational dynamics that is itself learnable: a structured object that an arbitrary downstream model can consume, that abstracts away the substrate from which the modes were derived, and that places different proteins, different conformations of the same protein, and in principle different biological scales into a single comparable space. The mode spectrum alone is a list of eigenvalues. The eigenvectors are tied to atomic coordinates and break under rotation, translation, and any change of basis. Neither object travels well across proteins or across scales.

Here we introduce frequency-space mechanics, a representational framework for biological dynamics in which the primary object of analysis is not a structure or a sequence but a graph over vibrational modes. Frequency-space mechanics is organised in three tiers: a framework, a state space, and an object. The framework itself, frequency-space mechanics, asserts that the relevant state space for biological function lives in the eigenstructure of the system's dynamics rather than in its instantaneous coordinates. The state space, which we call the latent mechanical space, is the abstract domain in which atomic positions have been projected out and only the harmonic relationships between vibrational modes remain. The object that lives in this space, one per protein per conformation, is the mechanical harmonics graph (MHG): a fully connected graph in which nodes are vibrational modes derived from molecular dynamics, edges are harmonic couplings weighted by octave alignment between mode frequencies, and node features encode the eigenvalue and the cumulative coupling weight at each mode. The MHG is coordinate-free, sequence-independent, and rotation- and translation-invariant by construction. Because its only inputs are eigenvalues and eigenvectors of a covariance matrix, the same construction applies unchanged to any system whose dynamics admit a tractable eigendecomposition, from femtosecond bond vibrations to slow collective motions in gene regulatory networks.

Working entirely within this framework and using no sequence information, we demonstrate three results. First, a graph neural network trained on static MHGs of 5,238 SwissProt proteins, under a strict 30% sequence-identity train-test split, predicts Gene Ontology molecular function terms across the ontology. Vibrational physics alone, with no evolutionary signal, encodes broad functional class. Second, applying Kuramoto entrainment to the harmonic coupling graph, formally a Hamiltonian operation over mode frequencies that pulls each node toward harmonic alignment with its neighbours, partitions protein function space into two regimes: a regime in which the static MHG is sufficient for prediction and a regime in which synchronisation dynamics specifically improve prediction for proteins whose function depends on collective conformational change. This partition is itself a biological finding. Third, on CLIC1, a fold- and function-switching chloride channel that was excluded from training by the homology filter, Kuramoto entrainment amplifies the channel-activity prediction signal 7.5-fold and the antioxidant prediction signal 2.4-fold, recovering both functional states of the protein from vibrational dynamics alone. CLIC1 occupies two completely separate branches of the GO molecular function ontology in its two folds, and the framework recovers both without ever seeing its sequence.

Beyond these results, frequency-space mechanics provides a representational substrate that is structurally compatible with quantum computing hardware. The Kuramoto entrainment step is a Hamiltonian minimisation over mode frequencies, a problem class that maps directly onto quantum annealing architectures[11], variational quantum eigensolvers[12], and quantum walks[13]. The MHG is therefore not only a learnable representation for classical graph neural networks but a formally specified problem instance that quantum solvers can address without modification to the rest of the pipeline. We return to this point in the Discussion.

# Results

## An overview of frequency-space mechanics

The frequency-space mechanics pipeline converts a system's dynamics into a coordinate-free graph representation through three stages (Figure 1). From an input structure we generate a short molecular dynamics trajectory and compute a force-displacement estimate of the Hessian, whose eigendecomposition yields the vibrational mode spectrum (Figure 1a). The pipeline's only physical input is a force-coordinate trajectory; all downstream constructs (mode spectrum, harmonic coupling graph, entrainment dynamics, classifier features) are derived from the estimated Hessian without further appeal to the underlying force generator. This single-input requirement is what makes the framework substrate-independent: forces from semi-empirical methods, density functional theory, classical molecular dynamics, or any future force generator enter the pipeline through the same interface, and the same construction extends to non-mechanical systems where state variables and their time derivatives are measurable. The mechanical harmonics graph is then constructed by placing one node per mode and connecting every pair of modes with an edge weighted by octave stiffness, an exponential kernel over the log-frequency ratio between the two modes that assigns maximal coupling to integer octave relationships and falls off smoothly with deviation from octave alignment, producing a fully connected graph in latent mechanical space (Figure 1b; Methods). Throughout this paper we use "octave" to denote a doubling of the Hessian eigenvalue $f = |\lambda|$, which plays the dimensional role of a squared frequency in the standard normal-mode formalism; in mass-weighted frequency $\omega = \sqrt{f}$, one octave in our sense corresponds to a $\sqrt{2}$ratio. This identification is deliberate: the MHG operates on the unweighted Hessian spectrum and inherits its scale, and the harmonic coupling kernel is defined on log-eigenvalue distances rather than log-frequency distances. The framework is otherwise unchanged under the substitution $f \rightarrow \omega$, and readers who prefer the $\omega$convention can apply it without modifying the pipeline. Five alternative entrainment operations were implemented, each producing an independent MHG from the static baseline by pulling mode frequencies toward harmonic alignment under a different dynamical rule. The Kuramoto synchronisation variant (V2) is used throughout this paper, while the other four variants serve as comparisons in the autoencoder and classifier analyses below (Figure 1c; Methods). The resulting MHG is the central object of analysis. Once constructed, it can be consumed by a graph neural network for function prediction, interrogated for entropy narrowing and hub emergence as signatures of conformational transition, or fed into a quantum solver as a Hamiltonian over mode frequencies. The applications in this paper exercise several of these consumers, but the pipeline through Figure 1c is the same in every case. The same three-stage construction applies unchanged to any system whose dynamics admit a tractable eigendecomposition, a property we return to in the Discussion.

## Training set and evaluation

We constructed MHGs for 5,238 proteins drawn from SwissProt under a filtering pipeline designed to maximise annotation quality and sequence diversity. Starting from 80,885 proteins with experimental Gene Ontology annotations, we required molecular function annotations at GO depth at least four, sequence length between 50 and 500 residues, mean AlphaFold pLDDT above 80, exclusion of any protein with greater than 30% sequence identity to the fold-switching validation protein CLIC1, and a final linclust pass at 50% identity to eliminate near-duplicates. The resulting set contained 2,897 unique molecular function terms after hierarchical propagation, of which 675 appeared in at least five proteins and formed the primary evaluation set. Proteins were split 70/30 into training and test sets with cluster-aware partitioning at 30% sequence identity, yielding 3,667 training proteins across 3,702 clusters with no cross-contamination between splits. Technical details of trajectory generation, Hessian estimation, MHG construction, entrainment variants, and classifier architectures are given in Methods.

## The frequency graph encodes learnable physical structure

Before testing whether the MHG predicts biological function, we first asked whether the physical organisation of the graph is internally consistent and learnable. We trained a masked group autoencoder (MLP, 10 to 16 to 4 to 16 to 10) on ten graph-level scalar features organised into three groups: edge coupling statistics (G1), node organisation statistics (G2), and global graph organisation metrics including frequency entropy and spectral gap (G3). For each variant we trained the autoencoder to reconstruct each masked group from the two unmasked groups, with no biological labels (Figure 2a).

On the V0 static baseline, reconstruction mean squared error was 0.045 for the edge coupling group, 0.161 for the global graph organisation group, and 0.525 for the node organisation group. The asymmetry is informative. Edge coupling statistics are almost entirely determined by node organisation and global graph organisation, and global graph organisation metrics are partially recoverable from the other two, but node organisation carries orthogonal information that cannot be reconstructed from the rest of the graph. Hub structure in frequency space is the most informationally independent feature of the MHG. Across entrainment variants, Kuramoto (V2) produced the most balanced reconstruction profile, while frequency-shifting variants V1 and V4 concentrated difficulty in the global graph organisation group, consistent with the fact that these variants move modes through the frequency spectrum while V0 and V2 largely preserve their physical positions (Figure 2b).

## Vibrational physics predicts molecular function without sequence

We next asked whether the MHG predicts molecular function directly. Two classifiers were trained on each entrainment variant: a multilayer perceptron over the ten graph-level scalar features, and a GATv2 graph neural network operating on the full node and edge feature sets of the MHG. Both were trained to predict the 675 primary GO molecular function terms under the 30% sequence-identity split, with no sequence information as input.

The graph neural network on static V0 MHGs achieved F_max of 0.039, hierarchical F1 of 0.106, and AUPR of 0.033 across the 675 terms, substantially exceeding the scalar MLP on identical inputs (F_max 0.029, HF1 0.023, AUPR 0.016; Figure 2c) and recovering most of the signal at the top of the GO hierarchy (depth-1 F_max 0.271, depth-2 F_max 0.159; Figure 2d). The absolute numbers are modest by the standards of sequence-based predictors, but they are obtained from vibrational physics alone on a strict homology split, which places them in a different reference frame. Depth-1 F_max above 0.25, without sequence, without homology, and from a 10-picosecond MD trajectory, establishes that broad molecular function class is encoded in vibrational dynamics.

## Kuramoto entrainment partitions protein function space

Globally, static physics outperformed all entrainment variants on the GNN classifier (V0 F_max 0.039, V1 0.037, V2 0.034), a finding we initially interpreted as evidence that entrainment smooths out biologically informative heterogeneity in the frequency graph. Per-term analysis revealed a more interesting structure. Comparing V2 Kuramoto against V0 static on each of the 675 terms individually, we found that the global average concealed a systematic partition.

Kuramoto entrainment produced the largest gains on terms related to receptor binding and signalling. G protein-coupled receptor binding (GO:0001664) improved by ΔF_max = +0.286 under V2, chemokine receptor binding by +0.160, cytokine binding by +0.121, and signalling receptor binding by +0.115. ATP-coupled transport terms formed a second V2-favoured cluster: ATPase-coupled transmembrane transporter activity, proton-transporting ATPase activity, and active ion transmembrane transporter activity all gained between +0.04 and +0.044. A third cluster comprised methyltransferase and kinase activities. These are precisely the functional categories in which collective conformational dynamics is the mechanism of function: induced fit in receptor-ligand recognition, conformational cycling in ATP-coupled transport, substrate channelling in methyltransferases. For static catalytic activities and broad binding categories, V0 raw harmonic structure was sufficient or superior (Figure 3a, 3b).

This partition is a biological finding, not an artefact of the classifier or the entrainment procedure. It states that vibrational physics predicts function through two distinct mechanisms: a direct path in which the raw eigenstructure encodes functional class, and an entrained path in which synchronisation dynamics reveal signal that is only visible once the graph has been allowed to relax toward its harmonic fixed points. Proteins fall into one camp or the other based on whether their function depends on collective motion.

## Direct interrogation of the harmonic coupling graph as a demonstration of framework interpretability

The partition finding raises a mechanistic question: why does Kuramoto entrainment specifically improve prediction for conformationally dynamic functions? To address this we interrogated the MHG directly on the CLIC1 fold-switching trajectory, using physics metrics extracted from the graph at every frame without any classifier in the loop. The 5,238 protein classification results presented above constitute the primary validation of the framework. The CLIC1 analyses presented in this and the following subsection demonstrate what the framework can reveal when interrogated directly on a single fold-switching protein, and serve as interpretability rather than as additional statistical validation.

We computed seven graph-level scalar metrics from Module 3 (frequency entropy, PhaseScore, node weight standard deviation, edge stiffness standard deviation, spectral gap, tensegrity index, and allosteric transfer efficiency) for both V0 static and V2 Kuramoto-entrained MHGs at every frame of the CLIC1 steered trajectory, in both forward (fold 1 to fold 2) and reverse directions. The results revealed four properties of the entrainment operation that we report as direct readouts of the harmonic coupling graph.

### *Kuramoto entrainment functions as a conformational amplifier.*

The static frequency graph is essentially insensitive to the conformational transition. Across all 500 forward trajectory frames, V0 metrics show coefficients of variation below 0.01, appearing as flat lines regardless of the protein's geometric state. Frequency entropy CV is 0.0060, PhaseScore CV is 0.0063, allosteric transfer efficiency (ATE), a graph-theoretic metric defined in Methods, CV is 0.0006. The static MHG of fold 1 and the static MHG of fold 2 are, by these scalar measures, nearly indistinguishable. Kuramoto entrainment applied to the same trajectory amplifies conformational sensitivity by 4 to 1217-fold depending on the metric (Figure 4a). ATE variance increases 1217-fold (CV 0.0006 to 0.730). Spectral gap variance increases 16-fold (CV 0.047 to 0.741). Frequency entropy variance increases 9-fold, PhaseScore 4-fold, and node weight standard deviation 2-fold. The forward and reverse trajectories, representing the enzyme-to-channel and channel-to-enzyme transitions respectively, are clearly distinguished by V2 metrics in ATE and spectral gap but indistinguishable under V0. These amplifications occur with no biological supervision and no classifier in the loop. They emerge purely from the entrainment dynamics applied to the physical force-displacement data.

### *Harmonic chain topology reorganises between fold states.*

Harmonic chain analysis at three trajectory frames reveals a clear topological change in the dominant coupling pathways through the MHG (Figure 4b). The fold 1 anchor frame shows two harmonic chains partitioning the vibrational mode space. The transition midpoint shows two chains in shifted proportions. The fold 2 anchor frame shows three chains with distinct gaps, a new harmonic coupling pathway that is not present in fold 1. The 2 to 3 chain reorganisation is consistent across stiffness thresholds of 0.5, 0.7, and 0.9 and is consistent with the formation of the ion channel pore as a new mechanical unit in fold 2. The chain topology is computed entirely from the V2 MHG with no structural annotation of the channel pore residues.

### *The mechanical centre shifts systematically across the trajectory.*

Per-residue participation in the dominant V2 vibrational mode shifts between key frames (Figure 4c). At the fold 1 anchor (frame 0), the dominant mode is concentrated on C-terminal residues 214, 224, 16, 206, and 65, distributed across the globular enzyme structure. At the frame of maximum ATE (frame 89, geometric RMSD 11.6 Angstroms from fold 1), participation has migrated to mid-protein residues 92, 94, 50, 54, and 61, forming a transient hub that coincides with peak allosteric coupling. By the fold 2 anchor (frame 499), the dominant mode is concentrated on residues 113, 169, 87, 130, and 134, in the putative channel-forming region. The mechanical centre therefore moves from C-terminus through a mid-protein hub to the channel region across the trajectory, identified entirely from vibrational dynamics without any structural annotation of which residues form the channel.

***Allosteric coupling reaches its maximum at intermediate geometric displacement.***

V2 Kuramoto entrainment identifies mechanically distinct states across the CLIC1 conformational trajectory, with allosteric transfer efficiency reaching its maximum at frame 89, a geometric displacement of 11.6 Angstroms from fold 1 and approximately 56% of the eventual maximum displacement of 20.8 Angstroms observed at frame 396. ATE peaks at intermediate geometry rather than at either fold endpoint or at the geometric midpoint of the trajectory, suggesting that the harmonic network passes through a state of maximum collective coupling at a specific point in the conformational pathway distinct from any geometric landmark. We do not interpret this as evidence about the physical folding mechanism, since the MD trajectory steers the system through a smoothed pathway rather than allowing it to find its own barrier crossing. We report the finding as a sensitivity analysis: V2 metrics distinguish geometrically intermediate states that V0 metrics cannot, and the maximum allosteric coupling identified by V2 does not coincide with the maximum geometric displacement.

Together these four findings provide a mechanistic interpretation of the partition result reported above. Kuramoto entrainment operates on the static MHG by converting subtle structural variation into large harmonic network reorganisations, including amplified scalar metric variance, topological changes in chain structure, and shifts in the mechanical centre. The classifier in the next subsection reads this transformed signal. The two-step mechanism, entrainment as amplifier followed by classifier as reader, is consistent with all four lines of evidence in the paper: hub emergence at the population level, physics metric amplification on the CLIC1 trajectory, GNN amplification of functional signal in the next subsection, and the per-term partition between static-sufficient and entrainment-required functional categories. We discuss the implications and limitations of this mechanistic claim in the Discussion.

## CLIC1: recovering fold- and function-switching from dynamics alone

Having shown that V2 entrainment amplifies physics metric variance across the CLIC1 trajectory without any classifier, we next asked whether this amplified signal translates into improved functional state recovery by the trained GNN. CLIC1 is a 226-residue protein that switches between a soluble glutaredoxin-like oxidoreductase fold with documented antioxidant activity (PDB 1K0N) and a membrane-inserted chloride channel fold (PDB 1RK4). The two folds occupy two completely separate branches of the GO molecular function ontology, and CLIC1 was excluded from training by the homology filter. We generated a 1000-frame steered MD transition trajectory spanning both folds, computed V0 and V2 MHGs at every frame, and extracted per-frame GNN predictions for channel activity and antioxidant activity GO terms.

Averaged over the trajectory, V2 Kuramoto entrainment amplified the channel activity prediction (GO:0015267) 7.5-fold relative to V0 (V0 mean 0.000254, V2 mean 0.001914; Figure 5a, 5b), amplified ligand-gated channel activity (GO:0015276) 4.0-fold, and amplified antioxidant activity (GO:0016209) 2.4-fold. The general binding term (GO:0005488) was unchanged or slightly reduced. CLIC1 falls within the ATP-coupled transporter category identified in the per-term analysis as V2-favoured, and the trajectory-level amplification on a protein excluded from training confirms that the partition discovered at the population level applies at the level of individual fold-switching proteins. Frequency-space mechanics, with no sequence input and no exposure to CLIC1 during training, recovered both functional states of the protein from vibrational dynamics alone (Figure 5c).

## Discussion

We have shown that protein molecular function can be predicted directly from the mechanical harmonics graph of a protein's vibrational dynamics, without sequence information, without homology, and under a strict 30% identity split. Four lines of evidence converge on a single mechanistic claim. Kuramoto entrainment of the MHG functions as a conformational amplifier, a nonlinear operation that maps narrow variation in the input Hessian onto wide variation in the harmonic coupling graph, and a graph neural network subsequently reads this transformed signal as functional information. At the population level, this amplification manifests as spontaneous hub emergence. At the single-protein level on the CLIC1 fold-switching trajectory, it manifests as 1217-fold amplification of allosteric transfer efficiency variance (ATE, a graph-theoretic metric), systematic shifts in the mechanical centre from C-terminus to channel region, and topological reorganisation of harmonic chains between fold states. At the classifier level, it manifests as 7.5-fold amplification of channel activity signal and 2.4-fold amplification of antioxidant activity signal on a protein excluded from training. At the per-term level, it manifests as a systematic partition of protein function space into categories where static physics suffices and categories where entrainment dynamics are required. Within the channel-related GO terms specifically, V2 improves prediction of channel activity, the collective ion-conduction mechanism, while V0 better predicts channel regulator activity, which depends on localised binding geometry that modulates channel function; this mechanistically coherent split supports the interpretation of the partition as a physical rather than statistical phenomenon. The two-step mechanism, entrainment as amplifier followed by classifier as reader, unifies these observations and identifies synchronisation dynamics on the harmonic coupling graph as the operation that exposes biological signal that is not visible in the static eigenstructure alone.

The direct MHG interrogation analysis on CLIC1 was motivated by a conceptual parallel to the HOMO-LUMO framework in quantum chemistry[14]. In that framework, stable molecular states correspond to occupied orbitals separated from reactive states by an energy gap that defines the reactivity window. In frequency-space mechanics, stable conformational states correspond to anchor modes (the fold 1 and fold 2 basins for CLIC1), and transition-prone states correspond to escape modes that are accessible only when coherent coupling focuses vibrational energy across a barrier in latent mechanical space. The CLIC1 data supports two components of this picture. The two fold states are distinguishable by V2 physics metrics and by harmonic chain topology, consistent with anchor modes as well-defined basins. The transient mid-protein hub identified at frame 89, where allosteric transfer efficiency reaches its maximum at an intermediate geometric displacement, is consistent with an escape mode that is not a geometric intermediate but a mechanical one. The third component of the analogy, a sharply localised entropy dip at the transition midpoint signifying the barrier crossing itself, was not resolved in the steered MD trajectory. The bias potential in MD steers the system through a smoothed pathway rather than allowing it to find its own barrier, and the monotonic variation of physics metrics across the trajectory is consistent with this. The entropy barrier signature predicted by the HOMO-LUMO analogy therefore remains a specific forward claim of the framework, testable under replica exchange sampling with explicit solvent and the Cys24 to Cys59 disulphide bond included (Future Directions).

The MHG under Kuramoto entrainment bears a structural resemblance to tensegrity, the architectural principle that distributed systems can achieve global stability through the balance of tension and compression across many weakly coupled elements. In real-space tensegrity, the load-bearing elements are compression struts stabilised by a network of tension cables, and the characteristic signature is load-bearing asymmetry: a small number of elements bear disproportionate stress while the network as a whole distributes force. In the MHG under V2 entrainment, harmonic coupling pulls modes toward octave alignment (the tension), while the physical eigenvalues resist runaway frequency shifts (the compression). The equilibrium reached is characterised by the same load-bearing asymmetry. A small number of hub modes accumulate disproportionate coupling weight while the population mean remains stable, a signature that reproduces the tensegrity organisational principle in eigenmode coordinates. We draw this parallel as a structural analogy with mathematical content rather than as a mechanistic identity. Where Ingber's tensegrity operates in real space on cytoskeletal elements[15], frequency-space mechanics operates in latent mechanical space on vibrational modes, which makes the framework substrate-independent and applicable to systems that have no real-space tensegrity at all.

The MHG is constructed as a fully connected graph with no edge stiffness threshold. This is a deliberate design choice rather than an oversight. A central premise of frequency-space mechanics is that global coordination in complex dynamical systems emerges from the integration of many individually weak interactions. The phenomenon is familiar from classical examples: pendulum clocks mounted on a shared wall gradually phase-lock through subtle mechanical transmission through the mounting surface[16], and a scattered audience clapping at different tempos spontaneously converges to a single rhythm through weak auditory coupling between neighbours[17]. In both cases, thresholding out the weak interactions would destroy the phenomenon. Our concern in constructing the MHG was that aggressive edge pruning might similarly discard biologically informative weak couplings before we had the opportunity to understand which ones matter. Fully connected construction preserves all harmonic coupling information and lets downstream learners decide which edges carry signal. It is almost certainly overkill for routine use at scale. Principled sparsification through learned cutoffs, attention mechanisms in the GNN, or spectral graph sparsification techniques that preserve essential topology while reducing edge count by one to two orders of magnitude is a natural area for future optimisation, particularly for extending the framework to larger systems and longer trajectories where the fully connected graph becomes computationally expensive.

A note on sampling. The 25-frame production window used here is below the threshold at which the displacement covariance $\langle \Delta x \cdot \Delta x^T \rangle$ is full-rank in the residue-resolved 3N coordinate system, and the Tikhonov regularisation specified in Methods sets the conditioning of the inversion. The estimated Hessian is therefore an effective Hessian over the subspace of modes populated within the trajectory rather than a full-rank reconstruction of the potential energy surface. The empirical adequacy of this projection for function prediction is established by the results above. Whether longer trajectories that recover the full-rank Hessian improve prediction differentially across functional categories is a question we leave to future work (Future Directions).

The Hessian computed in this paper is one specific choice. A 10-picosecond classical MD trajectory captures fast vibrational dynamics but is blind to slower collective motions that unfold over nanoseconds to microseconds and beyond, and equally blind to sub-picosecond quantum vibrational structure. The frequency-space mechanics framework places no constraint on the timescale at which the covariance matrix is computed. Quantum Hessians from semi-empirical or DFT calculations resolve bond-level vibrational detail; microsecond all-atom trajectories capture allosteric transitions and domain motions; ensemble-averaged Hessians over biologically relevant conformational states capture effective mechanics of the equilibrium ensemble. Most importantly, the harmonic coupling structure of the MHG provides a natural substrate for integrating Hessians computed at different timescales into a single multi-scale representation. Modes from a quantum Hessian occupy the highest frequency band; modes from picosecond classical trajectories occupy the middle band; modes from microsecond enhanced sampling occupy the low-frequency band. Because the octave stiffness kernel couples modes by frequency ratio rather than by source, modes from different timescales integrate into a single graph through their harmonic relationships across bands. This is a specific and distinctive prediction of frequency-space mechanics: harmonic coupling is the integration layer, and cross-scale representation does not require a new formalism, only the application of the existing MHG construction to Hessians sourced from multiple timescales.

Of the five entrainment variants implemented here, one (V4, the quantum tunneling surrogate) is deliberately included as a classical proxy for quantum annealing dynamics. V4 is not treated as a standalone result in this paper. Its purpose is to establish the formal problem structure that a genuine quantum solver would occupy: a Hamiltonian minimisation over mode frequencies on the spring-energy landscape of the MHG, compatible with quantum annealing architectures, variational quantum eigensolvers, and quantum walk algorithms. Quantum annealing in particular is among the quantum architectures closest to commercial deployment[18], with the principal limit on its use being the relatively narrow set of real problems that can be cast as the optimisation problems it is built to address. By formulating biological dynamics as harmonic coupling minimisation on the MHG, frequency-space mechanics expands the set of biologically meaningful problems that map naturally onto quantum annealing hardware, converting protein function prediction, conformational sampling, and the entrainment step itself into problem instances that quantum annealing solvers can address without architectural modification. A companion study currently in preparation replaces the classical V4 surrogate with D-Wave hybrid and native QPU solvers on the same MHG instances, using the binary quadratic model files generated by the V4 implementation as a shared artefact to enable direct three-way comparison between classical tunneling simulated annealing, hybrid quantum-classical solvers, and native QPU annealing. The frequency-space mechanics framework is therefore structured to accept quantum hardware as a drop-in replacement for a specific module of the entrainment step, without modification to any other part of the pipeline. This is a formally specified interface rather than an analogy, and its test is a separate experimental question rather than a conceptual extension.

The covariance source abstraction in Module 1 is not merely an architectural convenience. It is a claim about where biological information lives. Because the MHG construction depends only on the eigendecomposition of a covariance matrix, the same pipeline applies unchanged to any system with a meaningful covariance structure: gene expression profiles across cell states, neural activity across cortical regions, equity returns across market sectors, ecological population dynamics across species. This sets up a specific testable hypothesis by analogy to large language models. LLMs pretrained on internet-scale text learn representations of syntactic and semantic motifs (attention patterns, hierarchical composition, long-range dependency) that transfer across tasks and genres because the underlying structure of language is shared across domains. We propose that covariance-derived harmonic coupling structures are the analogue: universal organisational patterns that appear wherever coupled dynamical systems reach equilibrium, independent of what the nodes physically represent. A foundation model pretrained on protein dynamics via masked MHG reconstruction would learn these motifs from the richest and most physically constrained source available, and could then be fine-tuned on downstream covariance-structured tasks where labels are scarce. The experiment is straightforward to specify and sits as a natural follow-on programme.

The current paper commits to specific parameterisation choices: octave stiffness with sharpness parameter beta equal to 5, Kuramoto synchronisation as the primary entrainment, ten scalar metrics in the autoencoder analysis, a GATv2 architecture with four layers and two attention heads. These choices were made to enable the analyses reported here, but the framework is a template. Alternative harmonic coupling kernels, entrainment parameters and new variants beyond the five implemented, and additional graph-level or per-mode metrics can each be explored without modifying the other modules. Beyond classification, the MHG is a learnable and differentiable representation of protein mechanics, which opens applications in steered molecular dynamics guided by MHG-derived objectives, generative protein design conditioned on target harmonic coupling structure, and inverse design of proteins with specified vibrational signatures. These directions are not extensions of the GNN classifier but independent consumers of the MHG as a representational object, and they inherit the coordinate-free and sequence-independent properties of the framework without additional architectural commitments.

The picosecond vibrational band addressed in this paper is one narrow segment of the full biological frequency continuum, which spans approximately 28 orders of magnitude from femtosecond bond vibrations through picosecond side-chain dynamics, nanosecond loop motions, microsecond domain motions, millisecond conformational transitions, second-scale enzymatic turnover, minute-to-hour signalling cascades, hour-to-day cell cycle dynamics, day-to-year developmental and physiological rhythms, and longer evolutionary timescales (Figure 6). The apparent diversity of biological processes across these scales conceals a deep mechanical unity: all of them admit covariance descriptions, and all of them therefore admit harmonic coupling graph representations within the same formal structure. Frequency-space mechanics is in this sense not merely a protein method but the formal infrastructure through which biological processes at all scales become comparable within a single representational space. The narrowness of the current paper's focus is a computational and experimental limit, not a limit of the framework. The mechanical continuity of biology from atoms to ecosystems, expressed as harmonic coupling structure on covariance matrices, is what makes the cross-scale and cross-substrate claims of this paper more than rhetorical. Each scale is one panel of a continuous object, and the framework provides the vocabulary in which they can be compared.

Taken together, frequency-space mechanics establishes a new representational domain for biological dynamics, distinct from sequence-based and structure-based representations including those derived from conformational ensembles, that is coordinate-free, sequence-independent, scale-invariant, and structurally compatible with both classical graph neural networks and quantum computing hardware. The specific results reported here, on protein molecular function prediction from vibrational dynamics under a strict homology split and with a fold- and function-switching case recovered from dynamics alone, serve as the first demonstration of a framework whose scope extends substantially beyond a single biological problem. The MHG is the object that carries this scope. Its construction depends only on the eigendecomposition of a covariance matrix, and that constraint is weak enough to admit systems from femtosecond quantum vibrations to gene regulatory networks to financial markets, and strong enough to guarantee that every such MHG lives in the same latent mechanical space as every other. What the present work establishes is that protein function, the most stringently selected biological property in the evolutionary record, is visible in this space. What remains to be tested is how much of the rest of biology is visible there too.

# Future Directions

The results presented here establish frequency-space mechanics as an empirically grounded representational framework, with the mechanical harmonics graph as its central object. Several directions for immediate and longer-term investigation follow directly from the framework's architecture, and from the specific observations and limitations reported above. We describe the most concrete of these below, grouped by theme: improvements to the physical simulation that grounds the Hessian, extensions of the framework to new representational regimes, applications to domains beyond protein function, and validation on additional fold-switching proteins.

**Simulation quality and timescale.** Current results use 10-picosecond molecular dynamics trajectories with 25 frames, a computationally tractable but limited sampling of the vibrational landscape. The force-displacement Hessian improves with longer trajectories and more frames, as slow conformational modes become accessible only at longer timescales. Future work will extend simulations to microsecond timescales and beyond, where allosteric transitions, domain motions, and the entropy barriers that precede conformational switching are fully captured. At microsecond resolution the frequency graph spans the complete vibrational spectrum from fast bond vibrations to slow collective modes, and is expected to substantially improve prediction of fine-grained molecular functions at GO depth 4 to 5, where 10-picosecond physics provides limited resolution. Quantum Hessians derived from semi-empirical or DFT calculations will be applied to selected proteins where bond-level vibrational detail is mechanistically relevant, particularly for enzymes whose catalytic mechanism depends on specific vibrational coupling between active site residues. The role of explicit solvent structure, ionic conditions, and bound cofactors in shaping the Hessian will be systematically characterised, as these factors are known to alter protein mechanical properties and are omitted from the current apo simulations. A specific test of the entrainment-as-selection interpretation proposed in Discussion is whether longer trajectories that recover the full-rank Hessian improve prediction differentially across functional categories: functions encoded in dynamically populated modes are expected to remain approximately unchanged under extended sampling, while functions whose mechanism requires modes outside the short-trajectory subspace are expected to improve. The mdCATH ensemble, which provides 5,398 protein domains sampled at multiple temperatures across millisecond-scale aggregate trajectory time, is a natural resource for this test[19].

**Conformational ensembles.** Single AlphaFold structures represent one conformation from a complex energy landscape. Conformational ensemble methods such as BioEMU[20] generate physically realistic distributions of accessible structures, enabling ensemble-averaged frequency graphs and prediction confidence distributions. Proteins with wide prediction distributions across an ensemble are expected to be conformationally heterogeneous, a signal of functional plasticity that is not available to sequence-based methods. Preliminary analysis on fold-switching proteins suggests that ensemble prediction distributions are bimodal, with peaks corresponding to the two functional states; systematic characterisation across a larger set of proteins will test whether ensemble-averaged MHGs capture functional plasticity as a distinctive signature.

**Improved simulation of the CLIC1 fold switch.** The CLIC1 results reported in this paper are based on a steered MD transition trajectory that actively drives the system between the two folds. The MD bias potential is necessary to sample the fold transition on tractable timescales, but the steering force applied during the trajectory may perturb the natural harmonic structure of the system and the monotonic variation of physics metrics across the trajectory is consistent with smoothed sampling rather than genuine barrier crossing. Future work on CLIC1 will include three extensions designed to address this limitation. First, more thorough molecular dynamics with explicit solvent rather than implicit or vacuum treatment, since solvent structure and ionic environment contribute directly to the effective force field and therefore to the Hessian. Second, inclusion of the Cys24 to Cys59 disulphide bond[21] under appropriate redox conditions, since the fold switch in CLIC1 is triggered by the redox state of this bond and its presence or absence meaningfully changes the mechanical context of the protein. Third, robust replica exchange molecular dynamics or analogous enhanced sampling methods to map the energy landscape of the fold transition without applying a directional steering force, which will allow MHG construction on unperturbed conformational states. Together these extensions will test the entropy barrier prediction of the HOMO-LUMO analogy directly, under sampling conditions that allow the system to find its own barrier crossing.

**Multiscale Hessian construction.** The force-displacement Hessian presented here is computed at a single resolution, using residue-averaged forces from atomistic MD at 10-picosecond timescales. A natural extension is a multiscale Hessian that integrates mechanical information across timescales and length scales simultaneously. At the quantum scale, semi-empirical and DFT calculations resolve bond-level vibrational modes in the femtosecond band. At the atomistic scale, side-chain rotations and loop motions contribute middle-frequency modes. At the domain scale, interdomain motions and allosteric communication contribute low-frequency collective modes. At the supramolecular scale, subunit dynamics in protein complexes contribute the slowest modes. A multiscale Hessian constructed from a hierarchy of calculations and simulations (quantum mechanical, short atomistic, intermediate coarse-grained, long enhanced-sampling) would capture the full mechanical spectrum of a protein in a single frequency graph, with the octave stiffness kernel of the MHG coupling modes across bands through their harmonic relationships. This is expected to substantially improve prediction of functions that depend on either very fast chemical mechanism or very slow collective dynamics, including catalytic bond rearrangement, allosteric regulation, mechanosensing, and molecular motor activity, which are currently underrepresented in the 10-picosecond frequency graph.

**Functional class partitioning and drug discovery.** The per-term analysis presented here reveals that Kuramoto entrainment specifically improves prediction for proteins whose function requires collective conformational dynamics, including receptor binding, ATP-coupled transport, and substrate channelling, while the static frequency graph better predicts localised catalytic activity. This partitioning of molecular function space into entrainment-sensitive and entrainment-insensitive classes warrants systematic investigation. Future work will characterise the full landscape of GO functional classes with respect to entrainment sensitivity, testing whether the distinction reflects a fundamental biological partition between functions that emerge from collective vibrational modes and those that depend on local structural features. Such a classification would have direct implications for drug target identification: proteins whose function is entrainment-sensitive may be more amenable to allosteric modulation than those whose function is encoded in static structural features, providing a physics-based prior for selecting targets where allosteric drug design is likely to succeed.

**Supramolecular assemblies and the mechanical ecology of the cell.** Solvent structure, ionic strength, pH, and bound metabolites, cofactors, and small molecules all contribute to the effective force field and therefore to the Hessian and the MHG. MD simulations of holo proteins, including bound cofactors, substrate analogues, allosteric modulators, and explicit solvent, will capture the full mechanical context of the assembly. At the supramolecular scale, protein complexes and membrane-embedded assemblies can be treated as extended mechanical systems within the same formalism, with the MHG constructed over the collective modes of the assembled system rather than of individual proteins. Future work will establish frequency-space mechanics as a tool for modelling the mechanical ecology of the living cell, where protein function is shaped by mechanical context as much as by intrinsic structure.

**Quantum annealing hardware.** The V4 quantum tunneling variant presented in this paper uses a classical simulated annealing surrogate. The frequency graph spring energy minimisation problem is formally specified as a Hamiltonian over mode frequencies that maps directly onto D-Wave quantum annealing architecture[22]. A companion study currently in preparation replaces the classical surrogate with D-Wave hybrid and native QPU solvers on the same MHG instances, testing whether genuine quantum tunneling finds lower-energy harmonic configurations and whether quantum-optimised entrainment improves biological function prediction beyond what classical optimisation achieves. The binary quadratic model files generated by V4 serve as shared input artefacts across all three solvers, enabling direct comparison between classical tunneling simulated annealing, hybrid quantum-classical solvers, and native QPU annealing on identical problem instances.

**Multi-omic and regulatory extensions.** The covariance source abstraction means the same pipeline applies to any system with a meaningful eigendecomposition. Gene regulatory networks, in which gene expression levels replace residue positions and transcriptional changes replace forces, yield normal modes that can be embedded into the MHG formalism without modification. Protein-protein interaction networks, metabolic flux covariance, and chromatin accessibility dynamics are equally amenable. Future work will test whether the entrainment variants that improve protein function prediction also improve prediction of gene regulatory state transitions, cellular phenotype switching, and metabolic reprogramming, examining whether synchronisation dynamics are a universal organising principle across biological scales rather than a protein-specific phenomenon.

**Foundation model and cross-domain transfer.** The current pipeline trains on 5,238 proteins with experimental GO annotations. A natural extension is a self-supervised foundation model pretrained on the full SwissProt database (approximately 500,000 proteins) via masked MHG reconstruction, with no biological labels required. Learned representations would be fine-tuned with lightweight task-specific heads for GO molecular function prediction, EC number assignment, binding site identification, and drug target classification. The strongest test of the framework's scale-independence claim is whether such a protein-pretrained encoder transfers to covariance-structured tasks outside biology. Equity return covariances across market sectors, neural activity covariances across cortical regions, and ecological population dynamics all admit the same eigendecomposition and the same MHG construction. If the universal motifs claim of frequency-space mechanics is correct, a protein-pretrained MHG encoder should provide a meaningful initialisation for fine-tuning on any of these domains without retraining the representational backbone. This is the operational test that converts scale-independence from an architectural property into an empirical claim.

**Second fold-switching validation.** A second fold-switching protein, RfaH, is a natural target for follow-up validation. RfaH undergoes local C-terminal domain refolding rather than whole-protein conformational change[23], providing a mechanistic contrast to CLIC1. Results of this validation, and of extensions to additional fold-switching proteins, will be reported separately.

# Methods

## Dataset construction and train-test split

Proteins were drawn from the reviewed portion of UniProtKB/Swiss-Prot release 2025_03, matched to predicted protein structures obtained from the AlphaFold Protein Structure Database v6[3,24,25]. The AFDB v6 Swiss-Prot bundle is synchronised to Swiss-Prot release 2025_03, comprises approximately 550,000 structures, and excludes sequences shorter than 16 or longer than 2700 residues under AFDB coverage constraints. Swiss-Prot entries are represented by the top-ranked AlphaFold 2 prediction across five models trained with different random seeds; for entries whose sequences have not changed, v6 inherits coordinates from v4 (November 2022) with updated UniProt annotations. From this starting set, entries were restricted to those with experimental Gene Ontology molecular function annotations under evidence codes IDA, IMP, IPI, IGI, IEP, HDA, HMP, HGI, HEP, IC, TAS, and PANTHER_CURATED, yielding 80,885 proteins as the starting point for filter application.

The following filters were applied in sequence: molecular function annotation required (50,190 remaining), GO depth at least 4 after hierarchical propagation (14,146 remaining), sequence length between 50 and 500 residues (9,159 remaining), mean AlphaFold pLDDT confidence above 80 (6,715 remaining), exclusion of any protein with greater than 30% sequence identity to CLIC1 (the fold-switching validation protein reported in Results) or to a second fold-switching reference protein reserved for ongoing validation work described in Future Directions (one protein excluded on the second reference, at 65% identity; 6,714 remaining), and a final MMseqs2[26] clustering pass at 30% sequence identity and 0.80 coverage with cluster-aware train-test splitting (5,238 remaining).

Gene Ontology molecular function annotations were retrieved from go-basic.obo (April 2026 release), yielding 10,123 non-obsolete MF terms. Each protein's leaf annotations were propagated to all ancestor terms at GO depth at least 1, producing a label matrix of 5,238 proteins by 2,897 unique terms after propagation. Terms appearing in at least 5 proteins (675 terms) formed the primary evaluation set used throughout Results. Terms appearing in at least 2 proteins (1,556 terms) formed an extended evaluation set, and singleton terms (1,341) were reserved as a zero-shot evaluation set. Mean labels per protein was 6.3 with label density 0.83%.

Proteins were partitioned into training and test sets under a cluster-aware 70/30 split using MMseqs2 sequence clustering at 30% identity and 0.80 coverage. Entire clusters were assigned to either train or test, yielding 3,667 training proteins across 3,298 clusters and 1,571 test proteins across 404 clusters. The total 3,702 clusters included 2,929 singleton clusters (79.1%).

No protein in the test set shared more than 30% sequence identity with any protein in the training set, verified by direct pairwise comparison.

## Structure preparation

Training set structures were AlphaFold 2 model_v6 predictions as described above, filtered to mean pLDDT greater than 80. Structures were prepared with OpenMM[27] Modeller by adding hydrogen atoms at pH 7.4 and removing any water molecules present in the source file. Missing residues were not modelled, and missing-atom additions to terminal or loop regions were suppressed to avoid introducing artefacts into the downstream force extraction.

CLIC1 fold-switching structures were prepared using PDBFixer from two experimentally determined conformations: fold 1 corresponds to PDB entry 1K0N chain A, the reduced soluble glutaredoxin-like oxidoreductase form, and fold 2 corresponds to PDB entry 1RK4 chain B, the oxidised membrane-competent chloride channel form. Missing atoms within resolved residues were identified and added by PDBFixer; missing residues and missing-atom additions to terminal or loop regions were suppressed. Hydrogens were added at pH 7.0. The prepared structures were energy-minimised for 5,000 steepest-descent steps in AMBER14 with OBC2 implicit solvent before entering the steered MD protocol.

## Classical molecular dynamics (training set)

All 5,238 training proteins were processed through an identical four-stage classical MD protocol implemented in OpenMM. The force field was AMBER14 all-atom (amber14-all.xml) with GBn2 implicit solvent (implicit/gbn2.xml). The choice of GBn2 for the training set reflects its improved accuracy for folded proteins relative to earlier Generalised Born variants. Nonbonded interactions used NoCutoff (standard for implicit solvent), bond constraints were applied to hydrogen bonds (HBonds constraint level), and hydrogen mass repartitioning at 1.5 amu was used to enable a 2 fs integration time step. Integration used a Langevin Middle integrator at 300 K with a friction coefficient of 1.0/ps.

The protocol consisted of four stages executed in sequence. Stage 0 was an L-BFGS energy minimisation of up to 5,000 iterations to remove steric clashes from the AlphaFold input. Stage 1 was a restrained heating phase in which Cα atoms were held by harmonic positional restraints at the minimised coordinates with spring constant $k = 1.0$ kcal/mol/Å², and the system was heated from 0 K to 300 K in ten equal temperature increments over 1 ps, followed by 1 ps at 300 K with restraints still in place. Stage 2 was a 3 ps unrestrained warmup at 300 K allowing side-chain and loop relaxation. Stage 3 was a 5 ps unrestrained production run at 300 K during which per-atom forces and positions were recorded every 100 integration steps (every 0.2 ps), yielding 25 frames per protein. Total simulation time per protein was approximately 10 ps; for each protein, only the 25 Stage 3 frames entered the downstream Hessian estimation pipeline.

Per-atom forces and positions were reduced to per-residue quantities within each frame before the frame was written to disk. Positions were reduced by taking the mean of all atom positions within each residue; forces were reduced by summing all atom forces within each residue, preserving the net force on each residue. The resulting per-residue force and position arrays were written in forceout_v1 TSV format with positions in nanometres and forces in kJ/mol/nm, producing one file per protein containing 25 frames of residue-averaged data. These files are the direct input to the Module 1 Hessian estimation procedure.

## CLIC1 steered MD transition trajectory

The CLIC1 fold-switching trajectory was generated using a steered MD lambda annealing protocol in OpenMM with the AMBER14 all-atom force field and OBC2 implicit solvent (implicit/obc2.xml). OBC2 was used here rather than GBn2 for consistency with a pre-existing implicit-solvent setup; the difference between the two implicit solvents does not affect the MHG pipeline, which consumes force-displacement covariance rather than absolute energies. Nonbonded interactions used NoCutoff, bond constraints applied to hydrogen bonds (HBonds), hydrogen mass repartitioning at 1.5 amu, and a Langevin Middle integrator at 300 K with friction 1.0/ps and time step 2 fs. The forward direction (fold 1 to fold 2, enzyme to channel) and reverse direction (fold 2 to fold 1, channel to enzyme) were run as independent simulations; no replica exchange was used.

Restraints guiding the fold transition were implemented as one-sided harmonic Cα–Cα contact restraints using OpenMM CustomBondForce. For each direction, the restrained contact set consisted of Cα–Cα pairs present in the target fold but absent in the starting fold, with Cα–Cα distance below 0.8 nm and sequence separation of at least 3 residues in the target structure. Contacts shared between the two folds were not restrained. The restraint potential per contact was $E = 0.5 \cdot k \cdot \lambda \cdot \mathrm{step}(r - r_0) \cdot (r - r_0)^2$ with spring constant $k = 100.0$ kJ/mol/nm², step() the Heaviside step function so that the restraint activates only when the current Cα–Cα distance $r$ exceeds the target $r_0$, and $r_0$ defined as the target-fold Cα–Cα distance plus a 0.1 nm buffer. $\lambda$ was the global annealing parameter, taking values 0 at the start of the trajectory and 1 at the final window. In the forward direction (fold 1 to fold 2), 158 contacts were shared between folds and 343 target-unique contacts were restrained. In the reverse direction (fold 2 to fold 1), 158 contacts were shared and 372 target-unique contacts were restrained out of 446 identified, with 74 excluded due to residue index mismatches between the coarse-grained contact map and the prepared all-atom structure.

The trajectory was run for 5,000,000 integration steps (10 ns total per direction) divided into 50 lambda windows of 100,000 steps each, with λ incremented linearly as $\lambda_i = i/49$ for i = 0 through 49 at the start of each window. During each window, per-residue forces and positions were sampled in windows of 25 samples taken every 50 integration steps (every 100 fs), yielding a 2.5 ps force window per recorded trajectory frame. Trajectory frames were written every 10,000 integration steps, giving 10 frames per lambda window and 500 frames total per direction (1000 frames across the combined forward and reverse trajectories used in Results). Per-atom forces and positions were reduced to per-residue quantities using the same convention as the training set protocol. Full all-atom PDB snapshots were saved at $\lambda = 0.00$, 0.25, 0.50, 0.75, and 1.00 for visualisation, but were not used in the MHG pipeline. Before the production annealing run, the prepared CLIC1 fold 1 and fold 2 structures were each energy-minimised for 5,000 steps (preparation minimisation) and a further 500 steps immediately before the production trajectory began.

## Force-displacement Hessian estimation

For each trajectory, the Hessian of the potential energy surface was estimated directly from force and displacement fluctuations sampled across the trajectory frames. Per-residue forces and positions were read from force output tables, with positions converted from nanometres to Angstroms and forces retained in kJ/mol/nm units. Force output was produced at the residue-averaged level by the upstream OpenMM extraction step, with no additional atom-to-residue aggregation in the pipeline.

Given per-frame position matrix x of shape (n_frames, 3N) and force matrix F of shape (n_frames, 3N) where N is the number of residues, the Hessian estimate was computed as follows. Displacements $\Delta x$ were defined as x minus the trajectory mean position. The force-displacement covariance was computed as $FC = (F^T \cdot \Delta x) / (n_{frames} - 1)$. The displacement auto-covariance was computed as $C = (\Delta x^T \cdot \Delta x) / (n_{frames} - 1)$. Tikhonov regularisation[28] was applied as $C \leftarrow C + \varepsilon \cdot I$ with $\varepsilon = 10^{-8} \cdot \mathrm{tr}(C) / 3N$ to ensure numerical invertibility. The Hessian was then computed as $H = -FC \cdot C^{-1}$, with numpy.linalg.inv used for the matrix inversion and a fallback to numpy.linalg.pinv on numerical failure.

The full 3N by 3N Hessian was diagonalised with numpy.linalg.eigh and eigenvalues were sorted by ascending absolute value. The six modes with smallest absolute eigenvalue were dropped as these correspond to the three translational and three rotational rigid-body degrees of freedom. No explicit eigenvalue threshold or rigid-body projection was applied. Absolute values were taken of the remaining $M = 3N - 6$ eigenvalues before downstream use, producing the vibrational mode spectrum. At 25 frames and a residue-resolved coordinate system, the displacement covariance $\langle \Delta x \cdot \Delta x^T \rangle$ is rank-deficient with respect to the full 3N coordinate space, and the Tikhonov regularisation described above sets the conditioning of the inversion. The regularised Hessian estimate is therefore an effective Hessian projected onto the subspace of modes populated within the trajectory, rather than a full-rank reconstruction of the underlying potential energy surface. The interpretation of this projection in relation to entrainment-based mode selection is discussed in Discussion.

## Mechanical harmonics graph construction

For each protein, a mechanical harmonics graph (MHG) was constructed with one node per vibrational mode (M nodes total) and an edge between every pair of modes ($M(M-1)/2$ undirected edges, stored in both directions as 2E directed edges). The graph is fully connected with no stiffness threshold applied during construction, preserving all harmonic coupling information.

The edge weight between modes i and j was defined as the octave stiffness kernel $k_{ij} = \exp(-\beta \cdot |\log_2(f_i / f_j) - \mathrm{round}(\log_2(f_i / f_j))|)$ with sharpness parameter $\beta = 5.0$, where $f_i$ and $f_j$ are the absolute-valued Hessian eigenvalues of modes i and j as defined in Results. The kernel takes its maximum value of 1 when $f_i$ and $f_j$ are related by an exact integer number of octaves including the unison $f_i = f_j$, and decreases exponentially with deviation from the nearest integer octave relationship, reaching a minimum of $\exp(-\beta/2) \approx 0.082$ at maximum deviation. The logarithmic frequency coordinate is necessary because harmonic coupling between oscillators is defined by frequency ratio rather than frequency difference, and the logarithmic metric preserves scale invariance across the vibrational spectrum. The kernel admits a spring-energy interpretation in which the MHG is viewed as a network of Hookean springs with spring constants $k_{ij}$ between nodes at log-frequency positions.

Node features were stored as an (M, 2) array containing the eigenvalue and the node weight. Node weight was defined as the cumulative sum of incident edge stiffnesses $w_i = \Sigma_j k_{ij}$, capturing each mode's total harmonic coupling strength to the rest of the graph. Edge features were stored as a (2E, 3) array in both directions containing stiffness $k_{ij}$, harmonic distance $|\log_2(f_i/f_j) - \mathrm{round}(\log_2(f_i/f_j))|$, and frequency ratio $|\log_2(f_i/f_j)|$.

## Entrainment variants

Five entrainment operations were implemented, each taking an input MHG and producing an independent output MHG with effective mode frequencies modified according to a variant-specific dynamical rule. All variants share the interface (graph → graph with effective frequencies) and can be substituted interchangeably in the downstream pipeline.

**V0 (static baseline):** Identity operation. Effective frequencies equal input frequencies.

**V1 (spring relaxation):** Fixed-iteration gradient descent on the spring energy over log-frequency coordinates. At each iteration, log-frequencies are updated as log_freqs ← log_freqs − $\eta$ · grad, where $\mathrm{grad}_i = \Sigma_j k_{ij} \cdot \mathrm{displacement}_{ij}$ and $\mathrm{displacement}_{ij} = \log_2(f_i) - \log_2(f_j) - \mathrm{round}(\log_2(f_i/f_j))$. Learning rate $\eta = 0.01$, maximum iterations 50, $\beta = 5.0$. No convergence-based early stopping.

**V2 (Kuramoto synchronisation):** Iterative application of the Kuramoto rule in which each mode's effective log-frequency is updated by pulling it toward harmonic alignment with all other modes, weighted by octave stiffness. At each iteration, with $diff_{ij} = log_freqs_i - log_freqs_j$ and $displacement_{ij} = diff_{ij} - round(diff_{ij})$, the coupling matrix $k_{ij} = \exp(-\beta \cdot |displacement_{ij}|)$ is computed with the diagonal zeroed, and the update $delta_i = -\kappa \cdot \Sigma_j k_{ij} \cdot \sin(2\pi \cdot displacement_{ij})$ is applied. Coupling strength $\kappa = 0.05$, 200 iterations, $\beta = 5.0$. To prevent runaway frequency shifts, log-frequencies are clamped at each iteration to within three octaves of the initial range using numpy.clip(log_freqs, log_freq_min − 3.0, log_freq_max + 3.0).

**V3 (soft assignment):** Node weights are replaced with softmax-weighted stiffness claim strength across harmonically aligned modes; frequencies themselves are not modified. Temperature parameter 1.0.

**V4 (quantum tunneling surrogate):** Classical simulated annealing over the spring-energy landscape with probabilistic uphill moves approximating quantum tunneling dynamics. 500 annealing steps with linear cooling schedule from initial temperature 1.0 to final temperature 0.01. Gradient step size 0.01. Tunneling events occur with Bernoulli probability 0.1 per step; when triggered, Gaussian noise of standard deviation equal to the current temperature is added to log-frequencies without Metropolis acceptance (uphill moves always accepted). Same ±3 octave clamping bounds as V2. Fixed RNG seed 42 for reproducibility.

## Module 3: graph-level scalar metrics

A set of graph-level scalar metrics was extracted from each MHG in a separate pass after graph construction and entrainment. Metrics were organised into three groups for the masked group autoencoder analysis. G1, edge coupling statistics: mean, standard deviation, skewness, and Gini coefficient of the edge stiffness distribution. G2, node organisation statistics: mean, standard deviation, skewness, and Gini coefficient of the node weight distribution. G3, global graph organisation metrics: frequency entropy, PhaseScore, tensegrity index, allosteric transfer efficiency, edge entropy, spectral gap, and degree assortativity.

Frequency entropy is the Shannon entropy of the normalised eigenvalue distribution: $H_{freq} = -\Sigma_i p_i \log(p_i)$ where $p_i = \lambda_i / \Sigma_j \lambda_j$ and $\lambda_i$ are the absolute-valued eigenvalues of the vibrational mode spectrum. A small regularisation term of $10^{-12}$ is added to normalisations and logarithm arguments throughout to prevent numerical underflow.

PhaseScore is a weighted composite of frequency entropy, node-weight entropy, and edge stiffness skewness: $PhaseScore = 0.4 \cdot H_{freq} + 0.4 \cdot H_{node} + 0.2 \cdot \gamma$, where $H_{node}$ is the Shannon entropy of the normalised node weight distribution and $\gamma$ is the skewness of the edge stiffness distribution. The three components are on different intrinsic scales — the entropies are bounded by log M while the skewness is dimensionless and typically O(1) — so the 0.4/0.4/0.2 weights are heuristic rather than derived. PhaseScore is used here as a single-number summary of harmonic organisation; component-wise normalisation to a common scale is a refinement reserved for future work.

Tensegrity index (TI) quantifies the balance of harmonic coupling weight between high-frequency and low-frequency modes: $TI = (W_{high} - W_{low}) / (W_{high} + W_{low})$, where $W_{high}$ and $W_{low}$ are the summed node weights of modes above and below the median eigenvalue respectively. TI takes values in (−1, 1), with positive values indicating coupling weight concentrated in high-frequency modes and negative values indicating low-frequency dominance.

Allosteric transfer efficiency (ATE) is a graph-theoretic transfer efficiency metric defined as the ratio of mean stiffness for long-range harmonic edges to mean stiffness for short-range edges: $ATE = \bar{k}_{long} / \bar{k}_{short}$, where $\bar{k}_{long}$ is the mean octave stiffness of mode pairs separated by more than three octaves ($|\log_2(f_i/f_j)| > 3$) and $\bar{k}_{short}$ is the mean stiffness of pairs within three octaves. ATE greater than 1 indicates that cross-register harmonic coupling is stronger than within-register coupling, a signature of long-range harmonic organisation. The name is motivated by the intuition that efficient propagation through a harmonic coupling network resembles allosteric signalling in proteins, but ATE is a property of the MHG rather than a direct measurement of biological allosteric communication. The 1217-fold amplification of ATE variance reported in Results for the CLIC1 trajectory under V2 entrainment refers to the coefficient of variation of this graph-theoretic quantity across the 500-frame trajectory (V0 CV = 0.0006, V2 CV = 0.730), not to changes in biological allosteric coupling.

Edge entropy (EE) is the Shannon entropy of the normalised edge stiffness distribution: $EE = -\Sigma_{ij} q_{ij} \log(q_{ij})$ where $q_{ij} = k_{ij} / \Sigma_{mn} k_{mn}$. High EE indicates coupling weight distributed broadly across edges; low EE indicates concentration in a few dominant edges.

Spectral gap is the algebraic connectivity of the weighted graph Laplacian, computed as the second smallest eigenvalue of L where L is constructed with stiffness as edge weights. Degree assortativity is the Pearson correlation of node degrees across edges, computed using the standard NetworkX implementation. Participation entropy is computed per mode as the Shannon entropy of the per-residue participation distribution; for mode m with eigenvector $u_m$ reshaped to $(n_{res}, 3)$, per-residue participation is $p_{mr} = (\Sigma_\alpha u^2_{mr\alpha}) / \Sigma_s \Sigma_\alpha u^2_{ms\alpha}$, and participation entropy is $H_m = -\Sigma_r p_{mr} \log(p_{mr})$.

## Masked group autoencoder

An MLP autoencoder was trained to reconstruct masked feature groups from unmasked ones. The encoder architecture was Linear(10 → 16) → LayerNorm(16) → ReLU → Linear(16 → 4), with a symmetric decoder of the same depth. For each entrainment variant, the model was trained with one group masked per training step, rotating across G1, G2, and G3. Loss was mean squared error computed only on the masked features. No biological labels were used at any stage of autoencoder training. Optimisation used Adam with learning rate $10^{-3}$ and weight decay $10^{-4}$ for 500 fixed epochs with full-batch updates. No dropout, early stopping, or additional regularisation beyond LayerNorm and weight decay.

## Classifiers and training

Two classifiers were trained on each entrainment variant. MLP scalar classifier: input was the ten graph-level scalar features. Architecture: Linear(10 → 64) → LayerNorm → ReLU → Dropout(0.1) → Linear(64 → 64) → LayerNorm → ReLU → Dropout(0.1) → Linear(64 → 675) → Sigmoid. Loss was binary cross-entropy. Optimiser: Adam with learning rate $10^{-3}$ and weight decay $10^{-4}$, with a CosineAnnealingLR schedule over $T_{max} = 200$. Batch size: full batch. Training ran for 200 fixed epochs with no early stopping.

GATv2[29] graph neural network classifier: input was the full MHG with node features (eigenvalue, node weight) of shape (M, 2) and edge features (stiffness, harmonic distance, frequency ratio) of shape (2E, 3). The encoder consisted of a node projection (Linear(2 → 64)), an edge projection (Linear(3 → 64)), and three stacked GATv2Conv layers with hidden dimension 64, 2 attention heads, edge dimension 64, concat heads with LayerNorm(128) and residual connections, followed by an output projection (Linear(128 → 64)). Graph-level embedding was obtained by global mean pooling over nodes, producing a 64-dimensional graph embedding. The classification head was Linear(64 → 128) → ReLU → Dropout(0.1) → Linear(128 → 675) with BCEWithLogitsLoss. Optimiser: Adam with learning rate $10^{-3}$ and weight decay $10^{-5}$. Scheduler: 10-epoch linear warmup followed by cosine decay with learning rate scaled by $0.5 \cdot (1 + \cos(\pi \cdot \text{progress}))$. Batch size was 1 graph per forward pass with gradient accumulation over 8 steps (effective batch size 8). Training ran with automatic mixed precision (torch.amp.autocast with GradScaler) and gradient clipping at max_norm 1.0. Early stopping with patience 20 epochs on validation loss. Maximum 200 epochs. Input features were globally standardised per variant using the Welford online algorithm applied to 200 training graphs.

Both classifiers used global normalisation of input features to prevent variant-specific normalisation from confounding the cross-variant comparison.

## Evaluation metrics and win counting

Evaluation used the 675 primary GO molecular function terms on the 1,571 test set proteins for each entrainment variant independently. $F_{max}$[30] was computed by sweeping a threshold $\tau$ over predicted probabilities, computing precision and recall at each $\tau$, computing $F1 = 2 \cdot \text{precision} \cdot \text{recall} / (\text{precision} + \text{recall})$, and retaining the maximum F1 across thresholds as F_max for the term. F_max was reported both globally (averaged across 675 terms) and stratified by GO ontology depth, where depth was defined as the shortest path from the term to the molecular function root (GO:0003674). Under this convention the three depth-1 terms of GO-MF are catalytic activity (GO:0003824), binding (GO:0005488), and antioxidant activity (GO:0016209). Hierarchical F1 (HF1) used the Clark and Radivojac[31] hierarchical evaluation in which precision and recall are computed over the ancestor-closed set of predicted terms weighted by information content. Area under the precision-recall curve (AUPR) was computed per term and averaged.

Per-term win counting for the partition analysis (Figure 3c). For each of the 675 primary terms, F_max was computed independently under V0 and V2. A term was counted as a V0 win if F_max(V0) − F_max(V2) > 0.001, as a V2 win if F_max(V2) − F_max(V0) > 0.001, and as tied otherwise. Terms with F_max of exactly zero under both variants (no prediction signal) were counted as tied. Across all 675 primary terms: V0 wins 114, V2 wins 68, tied 493. Win counts stratified by GO depth are shown in Figure 3c and summarised below:

| **GO depth** | **V0 wins** | **V2 wins** | **Tied** | **Total** | |
|---|---|---|---|---|---|
| 1 | 5 | 1 | 3 | 9 | |
| 2 | 16 | 7 | 6 | 29 | |
| 3 | 35 | 24 | 74 | 133 | |
| 4 | 40 | 22 | 157 | 219 | |
| 5 | 13 | 11 | 166 | 190 | |
| 6+ | 5 | 3 | 87 | 95 | |
| Total | 114 | 68 | 493 | 675 | |

The growth of the tied category at deeper GO levels reflects shrinking per-term sample sizes at those levels: deeper terms have fewer annotated proteins, and terms with very few positive proteins produce zero F_max under both variants which counts as tied.

## Software and computational resources

All pipeline code was implemented in Python 3.10.12. Core dependencies included PyTorch [32] 2.11.0+cu126 (CUDA 12.6), PyTorch Geometric 2.7.0[33], NetworkX 3.4.2, NumPy 2.2.6, SciPy 1.15.3, scikit-learn 1.7.2, pandas 2.3.3, matplotlib 3.10.8, UMAP-learn 0.5.11, and MDAnalysis 2.9.0. The training set MD pipeline ran inside a Docker container based on condaforge/mambaforge:latest with OpenMM installed from conda-forge with cuda-version=12 and the latest PDBFixer. The CLIC1 steered MD pipeline used the same OpenMM conda-forge build with cuda-version=12 on a dedicated GPU instance. Harmonic chain detection used HDBSCAN clustering[34]. Classifier training was performed on a single NVIDIA Tesla T4 GPU.

## Data and code availability

Data and code generated in this study will be deposited to public repositories upon peer-reviewed publication.

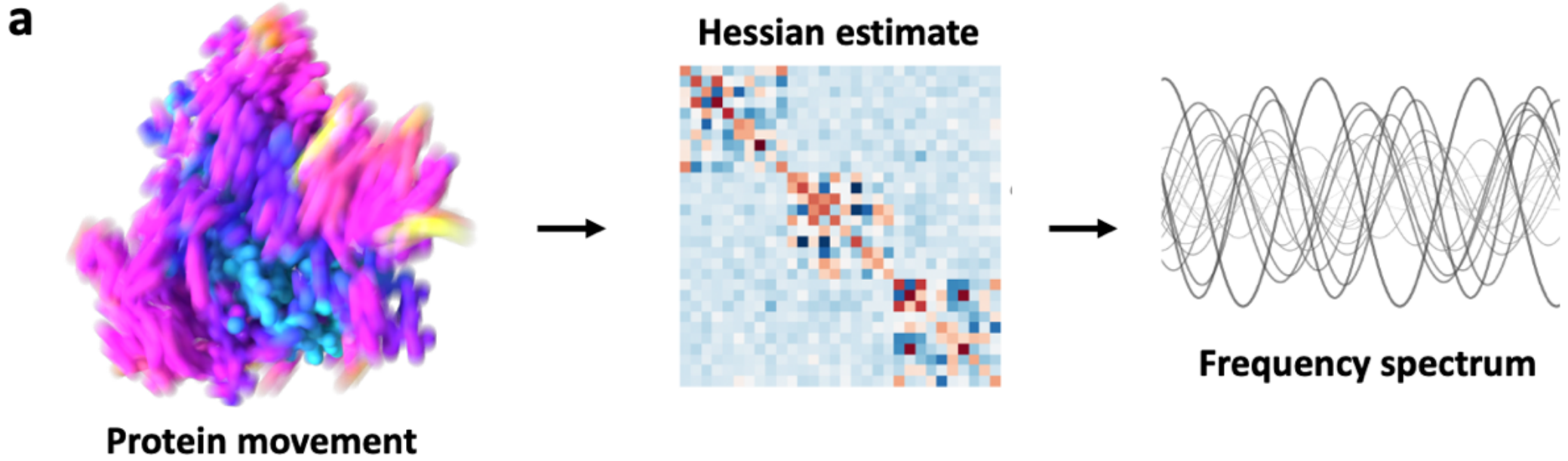

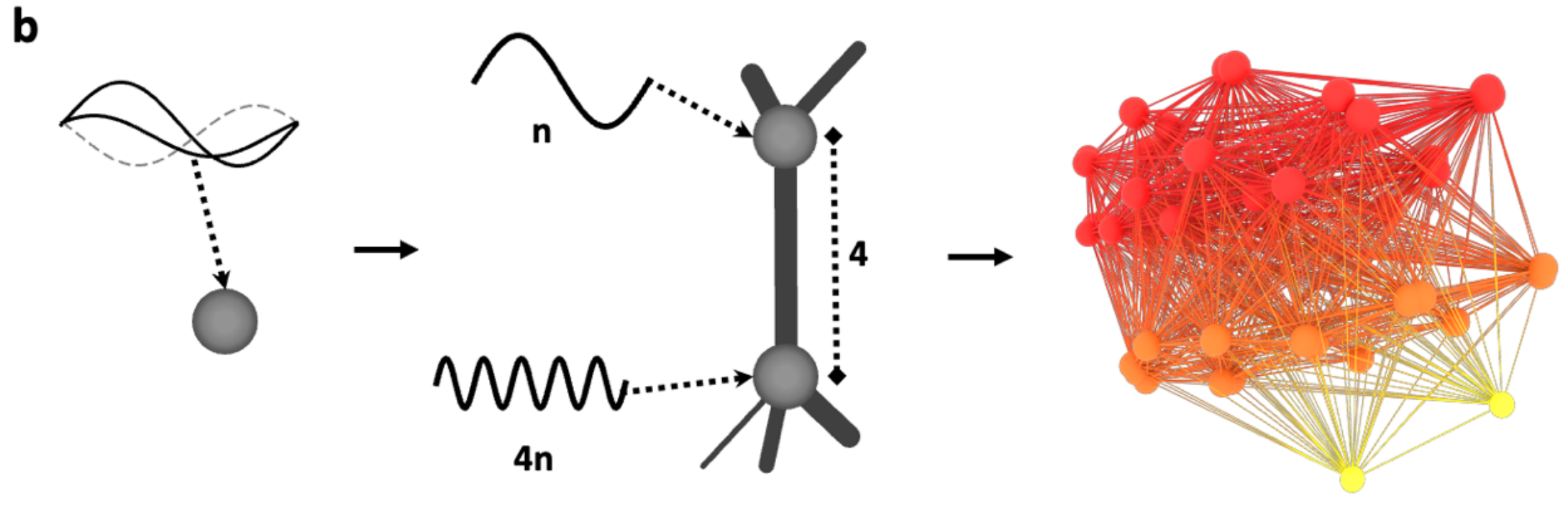

Harmonic graph construction

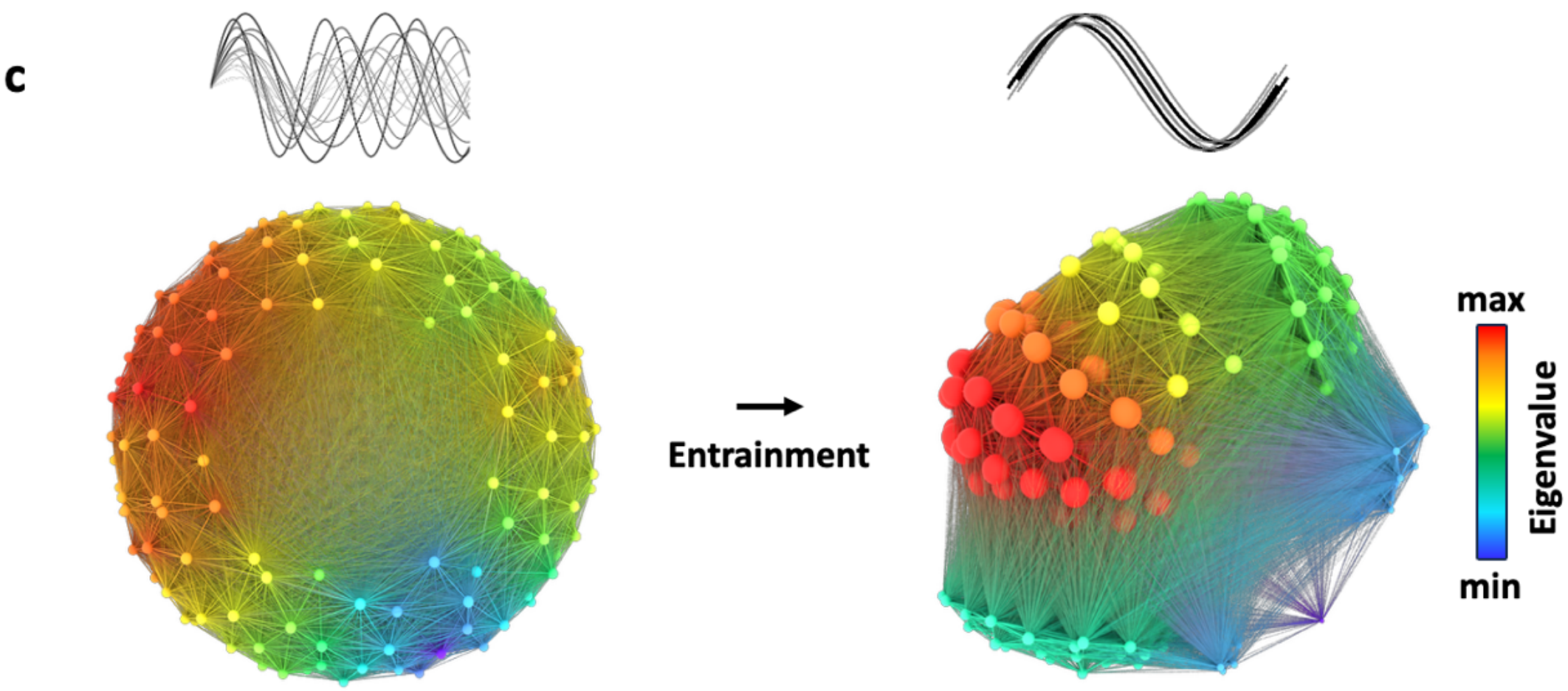

Fig 1.

**Figure 1. Constructing a mechanical harmonics graph in latent mechanical space.** (a) From an input structure, a short molecular dynamics trajectory is generated and used to compute a force-displacement estimate of the Hessian. Eigen decomposition yields the vibrational mode spectrum, with the six trivial modes removed. (b) The mechanical harmonics graph (MHG) is constructed by placing one node per vibrational mode and connecting every pair of modes with an edge weighted by octave stiffness. The graph is fully connected, sequence-independent, and coordinate-free. Node size encodes node weight (cumulative incident stiffness); edge thickness encodes pairwise stiffness, computed as an exponential kernel over the log-frequency ratio between two modes with maximum stiffness at integer octave relationships. (c) An entrainment operation is optionally applied to pull mode frequencies toward harmonic alignment. Five alternative variants are implemented, each producing an independent MHG from the static baseline; Kuramoto synchronization (V2) is used throughout this paper and produces the spontaneous emergence of high-weight hub modes shown here. The MHG, static or entrained, is the representational object that downstream applications consume. *Visualization note: the MHG inhabits a latent mechanical space with no real-space coordinates. For visualization purposes only, panels (b) and (c) project the graph back into a Cartesian layout via spring-energy minimization. This projection is not part of the pipeline; downstream applications consume the MHG directly as a graph object with node and edge features, with no spatial embedding at any stage.*

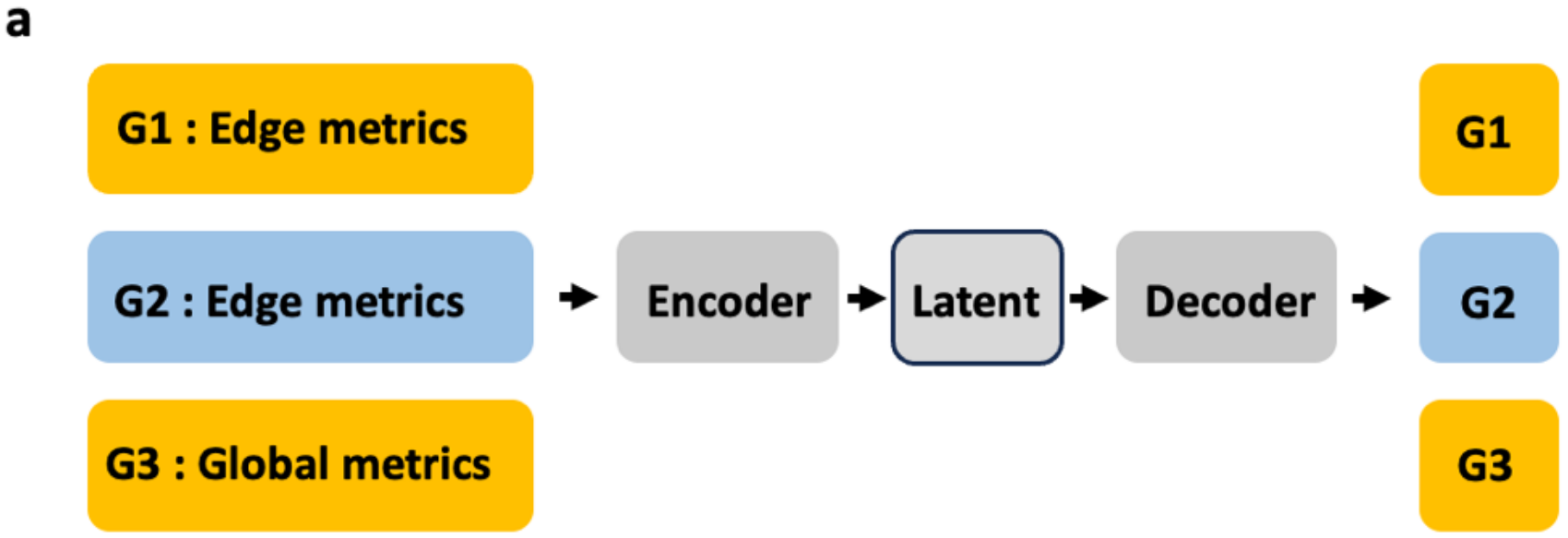
a
G1 : Edge metrics
G2 : Edge metrics
G3 : Global metrics
Encoder
Latent
Decoder
G1
G2
G3

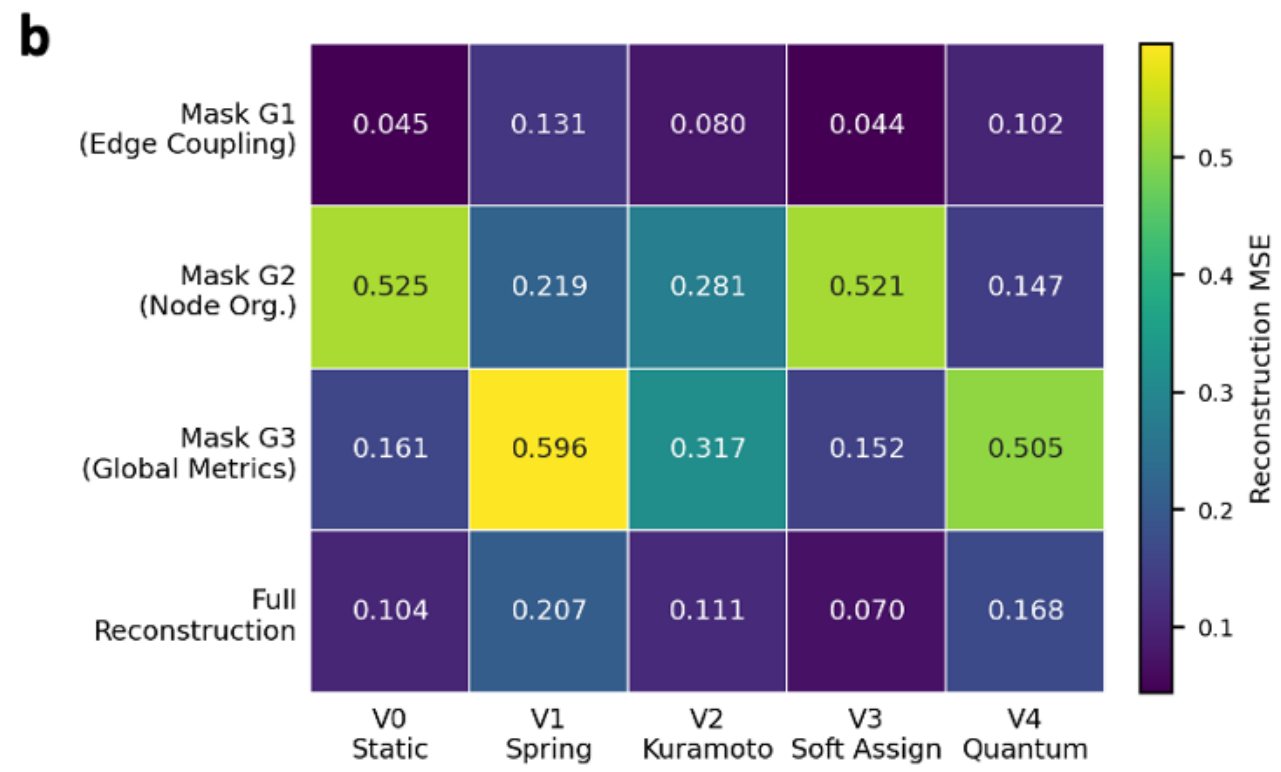
b
Mask G1 (Edge Coupling)
Mask G2 (Node Org.)
Mask G3 (Global Metrics)
Full Reconstruction
0.045 0.131 0.080 0.044 0.102
0.525 0.219 0.281 0.521 0.147
0.161 0.596 0.317 0.152 0.505
0.104 0.207 0.111 0.070 0.168
V0 Static
V1 Spring
V2 Kuramoto
V3 Soft Assign
V4 Quantum
Reconstruction MSE

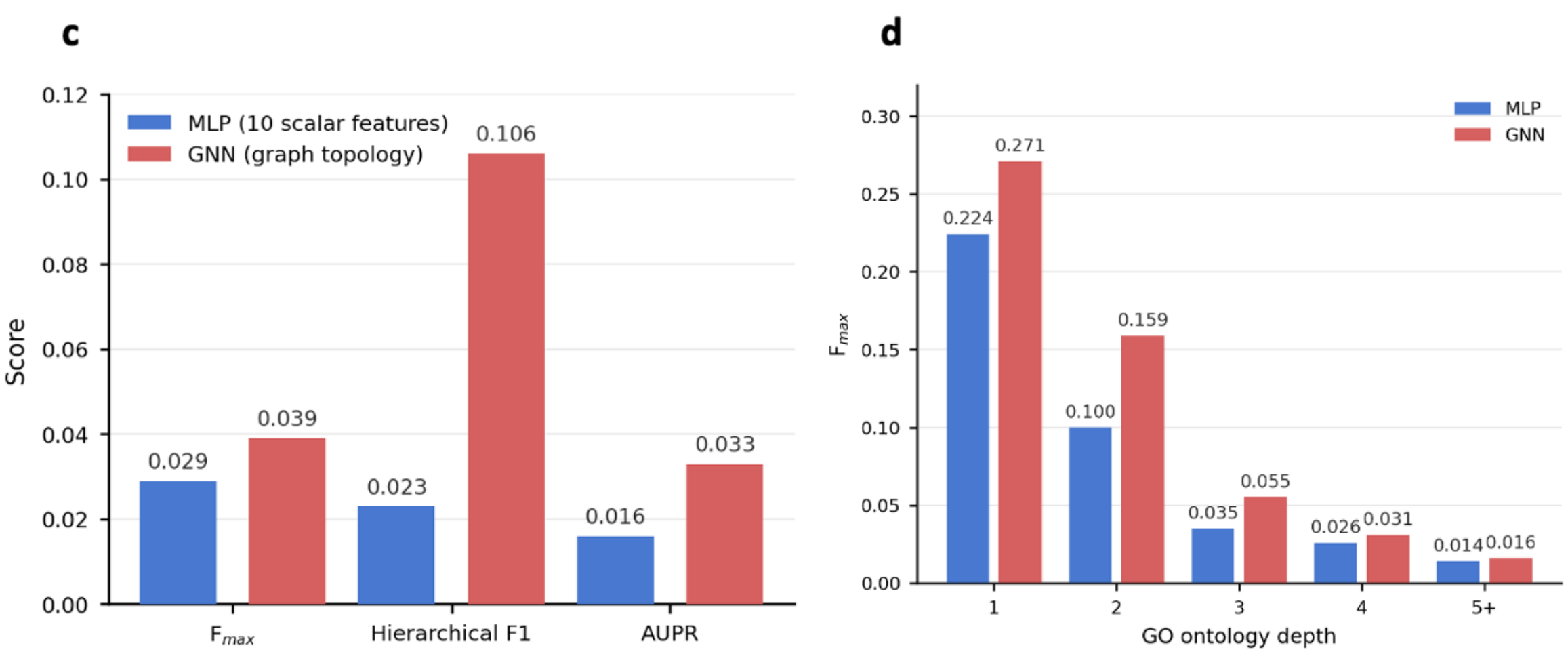
c
Score
MLP (10 scalar features)
GNN (graph topology)
0.029
0.039
0.023
0.106
0.016
0.033
$F_{max}$
Hierarchical F1
AUPR
d
$F_{max}$
MLP
GNN
0.224
0.271
0.100
0.159
0.035
0.055
0.026
0.031
0.014
0.016
GO ontology depth

**Fig 2.**

**Figure 2. The mechanical harmonics graph encodes learnable physical structure that predicts molecular function from vibrational dynamics alone.** (a) Masked group autoencoder architecture. Ten graph-level scalar features extracted from the MHG are organized into three groups (G1 edge coupling statistics, G2 node organization statistics, G3 global graph organization metrics including frequency entropy and spectral gap). The autoencoder is trained to reconstruct each masked group from the two unmasked groups, with no biological labels, across all five entrainment variants. (b) Reconstruction mean squared error (MSE) per masked group, per entrainment variant. Lower MSE indicates a feature group that is more predictable from the other two. The G2 node organisation row shows uniformly high MSE across variants, identifying hub structure as the most informationally independent feature of the MHG. (c) GNN versus MLP classifier comparison on V0 static MHGs across 675 primary GO molecular function terms. The GNN operating on full graph topology substantially exceeds the scalar MLP on identical inputs across F_max, hierarchical F1, and AUPR. (d) GNN classifier F_max stratified by GO ontology depth. Recovery of broad functional class at depths 1 and 2 is obtained from vibrational physics alone under a strict 30% sequence-identity train-test split, with no sequence information as input.

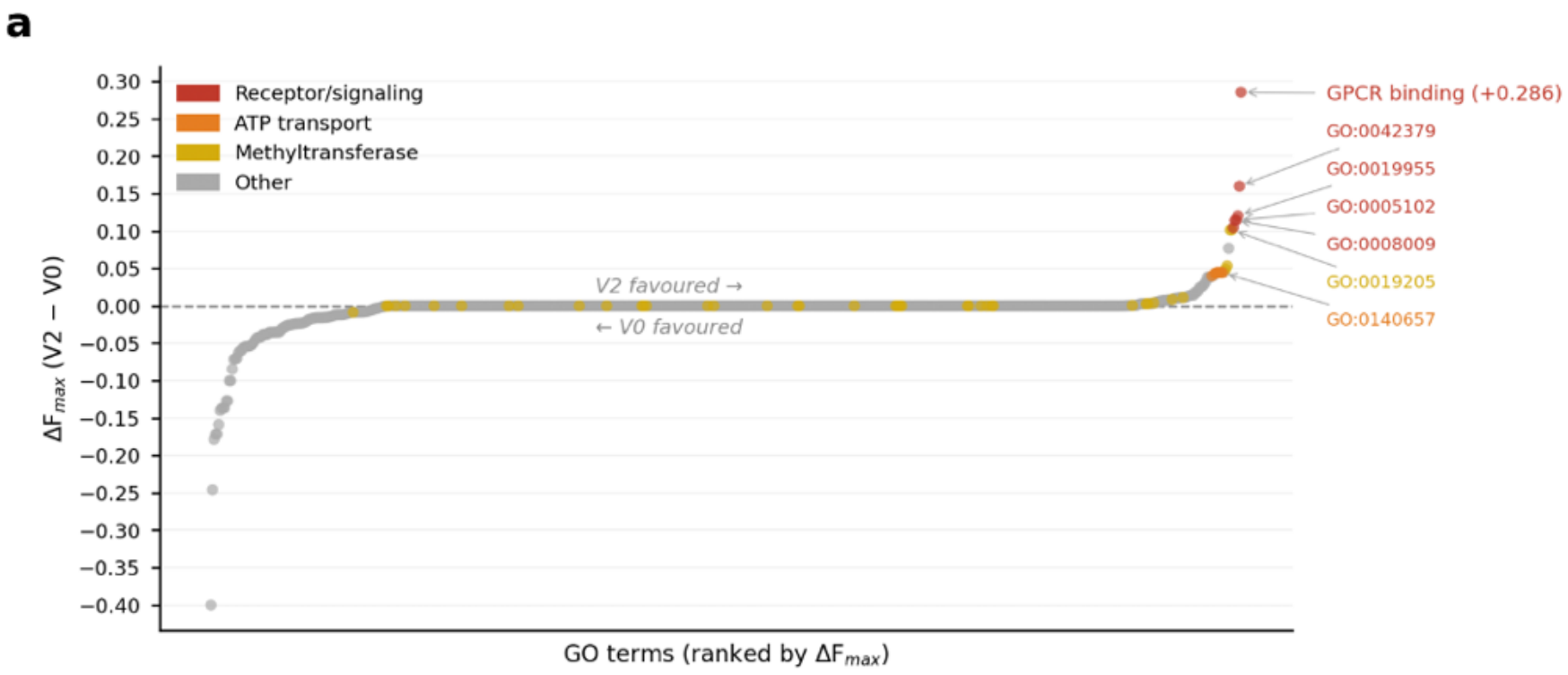
a
Receptor/signaling
ATP transport
Methyltransferase
Other
GPCR binding (+0.286)
GO:0042379
GO:0019955
GO:0005102
GO:0008009
GO:0019205
GO:0140657
V2 favoured →
← V0 favoured
ΔF_max (V2 − V0)
GO terms (ranked by ΔF_max)
0.30
0.25
0.20
0.15
0.10
0.05
0.00
−0.05
−0.10
−0.15
−0.20
−0.25
−0.30
−0.35
−0.40

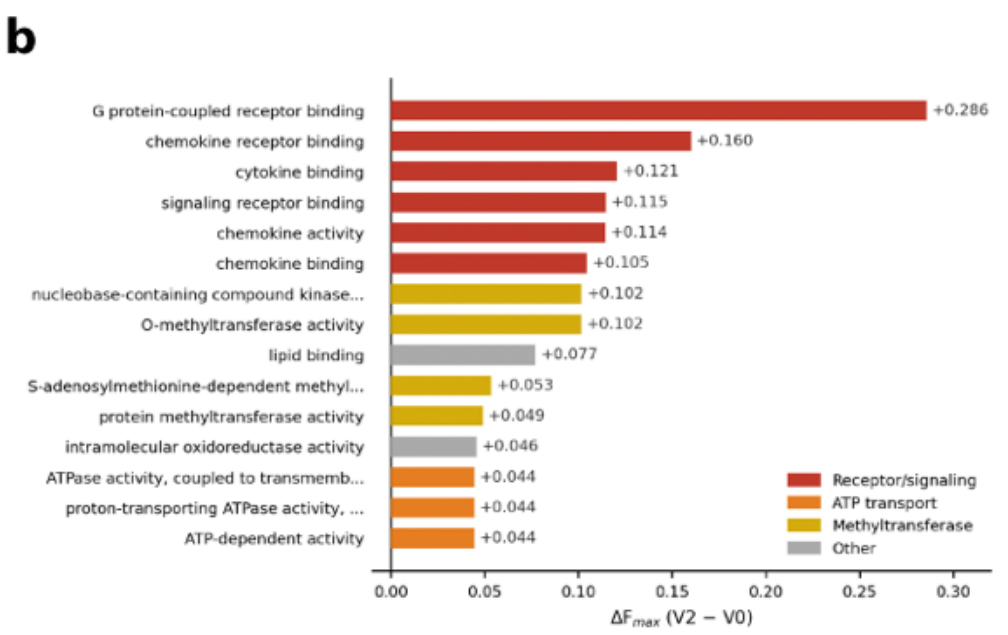
b
G protein-coupled receptor binding
chemokine receptor binding
cytokine binding
signaling receptor binding
chemokine activity
chemokine binding
nucleobase-containing compound kinase...
O-methyltransferase activity
lipid binding
S-adenosylmethionine-dependent methyl...
protein methyltransferase activity
intramolecular oxidoreductase activity
ATPase activity, coupled to transmemb...
proton-transporting ATPase activity, ...
ATP-dependent activity
+0.286
+0.160
+0.121
+0.115
+0.114
+0.105
+0.102
+0.102
+0.077
+0.053
+0.049
+0.046
+0.044
+0.044
+0.044
Receptor/signaling
ATP transport
Methyltransferase
Other
0.00
0.05
0.10
0.15
0.20
0.25
0.30
ΔF_max (V2 − V0)

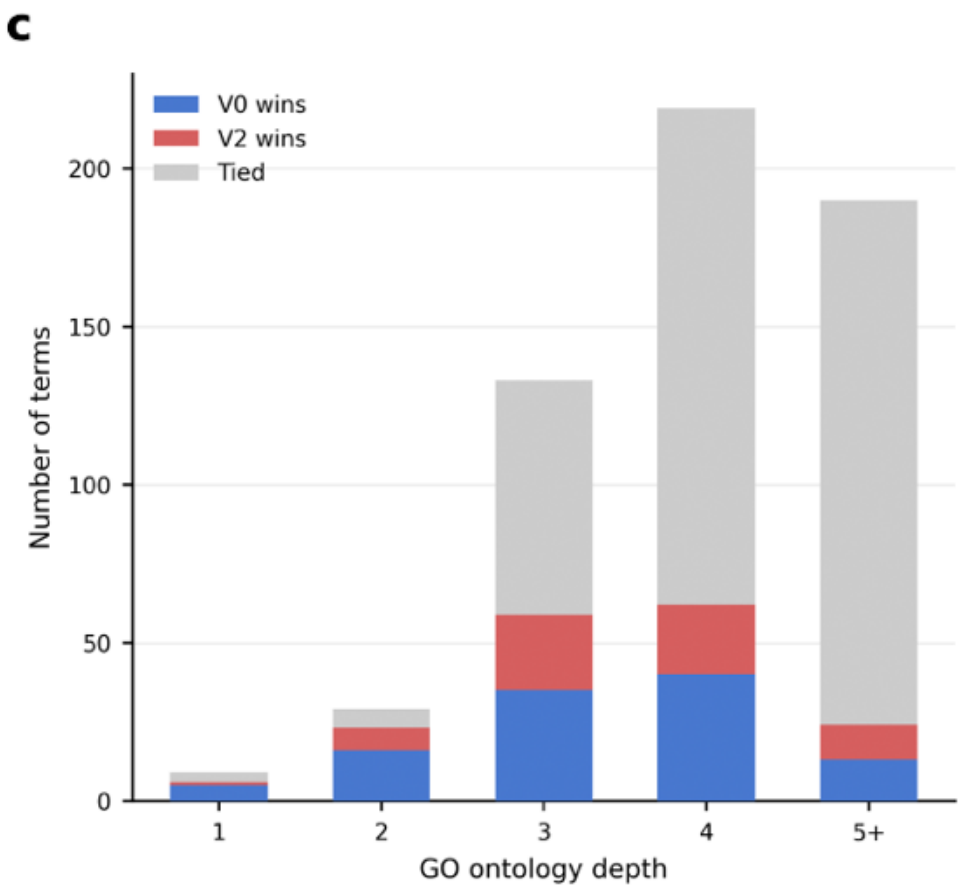
c
V0 wins
V2 wins
Tied
Number of terms
0
50
100
150
200
1
2
3
4
5+
GO ontology depth

**Fig 3.**

**Figure 3. Kuramoto entrainment partitions protein function space.** (a) Per-term V2 minus V0 $\Delta F_max$ across all 675 primary GO molecular function terms. Each point is one GO term; positive values indicate terms where Kuramoto entrainment improves prediction over the static baseline, negative values indicate terms where the static MHG performs better. Receptor binding and signalling terms (red), ATP-coupled transport terms (orange), and methyltransferase and kinase terms (gold) cluster in the V2-favoured regime; localized catalytic activities and broad binding categories cluster in the V0-favoured regime. (b) Top V2-favoured GO terms ranked by $\Delta F_max$. G protein-coupled receptor binding (GO:0001664) shows the largest gain at $\Delta F_max = +0.286$, with the receptor binding cluster occupying the top of the ranking. (c) Win counts by GO depth, V0 versus V2 across the 675 primary terms. The partition pattern shifts with depth: V0 dominates at shallow depths where broad functional categories favour static physics, while V2 wins increase at deeper, more specific functional terms. The number of tied terms grows substantially with depth as per-term sample sizes shrink.

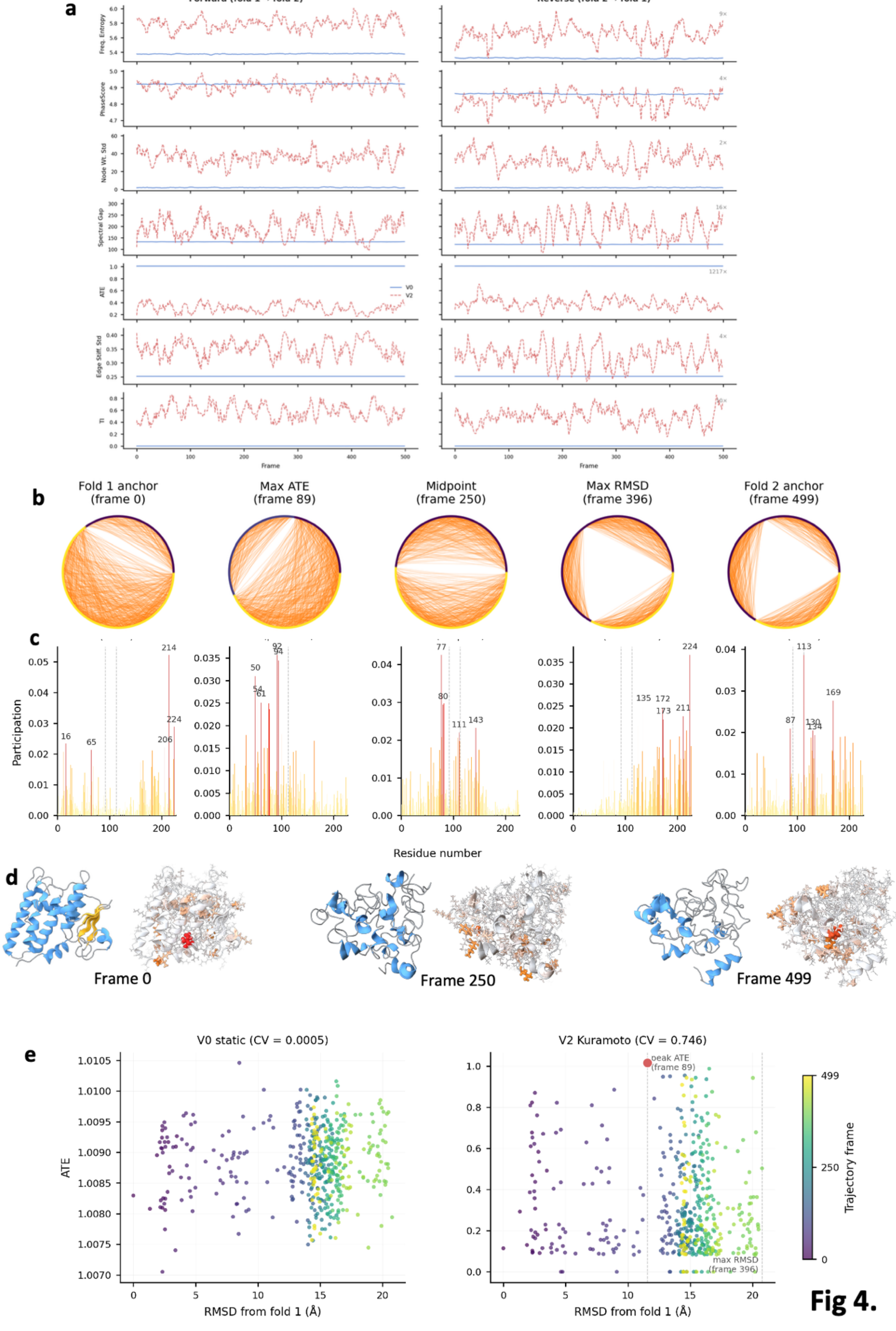

a
Forward (fold 1 → fold 2)
Reverse (fold 2 → fold 1)
Freq. Entropy
PhaseScore
Node Wt. Std
Spectral Gap
ATE
Edge Stiff. Std
TI
V0
V2
Frame
b
Fold 1 anchor (frame 0)
Max ATE (frame 89)
Midpoint (frame 250)
Max RMSD (frame 396)
Fold 2 anchor (frame 499)
c
Participation
Residue number
d
Frame 0
Frame 250
Frame 499
e
V0 static (CV = 0.0005)
V2 Kuramoto (CV = 0.746)
ATE
RMSD from fold 1 (Å)
peak ATE (frame 89)
max RMSD (frame 396)
Trajectory frame
Fig 4.

**Figure 4. Direct interrogation of the mechanical harmonics graph across the CLIC1 fold-switching trajectory.** (a) V0 static and V2 Kuramoto-entrained physics metrics across the CLIC1 sampling of the fold 1 to fold 2 conformational pathway, shown as forward (500 frames, fold 1 to fold 2) and reverse (500 frames, fold 2 to fold 1) independent simulations. Each row shows one scalar metric (frequency entropy, PhaseScore, node weight standard deviation, spectral gap, allosteric transfer efficiency (ATE), frequency entropy standard deviation, edge stiffness standard deviation) with forward and reverse trajectories as side-by-side subpanels sharing a common y-axis per row. V0 metrics (solid lines) show coefficients of variation below 0.01, appearing as flat lines regardless of conformational state. V2 metrics (dashed lines) show conformational sensitivity amplified 4 to 1217-fold depending on the metric, with the amplified dynamic range consistent across forward and reverse directions. (b) Harmonic chain topology at five trajectory frames under V2 entrainment, visualized as chord diagrams. Vibrational modes are shown as small circular nodes arranged around the perimeter in frequency order. Chains are sequences of strongly coupled modes identified at stiffness threshold 0.7 with minimum chain length 3; edges between mode pairs within a chain are drawn as curved Bezier chords across the interior of the circle, with all modes and edges belonging to the same chain sharing a common colour. The fold 1 anchor shows two harmonic chains, visible as two distinct chord populations. The fold 2 anchor shows three chains, with the emergence of a new chord population (purple) corresponding to a new harmonic coupling pathway absent in fold 1. The 2 to 3 chain reorganization is stable across stiffness thresholds of 0.5, 0.7, and 0.9. (c) Per-residue participation in the dominant V2 vibrational mode at five key trajectory frames (frame 0 fold 1 anchor, frame 89 peak ATE intermediate, frame 250 midpoint, frame 396 maximum geometric displacement, frame 499 fold 2 anchor). The mechanical centre shifts from C-terminal residues at fold 1, through mid-protein residues at peak ATE, to channel-region residues at fold 2, identified entirely from vibrational dynamics without structural annotation. Dashed vertical lines at residues 92 and 113 mark the peak-ATE and fold 2 dominant residues respectively. (d) CLIC1 structures at the fold 1 anchor, mid-point, and fold 2 anchor frames, colored by per-residue participation in the dominant V2 mode. Hot colours (red) indicate residues with highest participation; cool colours (white) indicate lowest participation. Renders indicating secondary structure are also provided with helix blue, and sheet yellow. (e) Relationship between allosteric transfer efficiency (ATE) and geometric displacement (RMSD from fold 1) across the trajectory. Each point is one trajectory frame, coloured by frame number. V0 and V2 are shown in separate subpanels. Under V0, ATE is essentially constant at 1.0 regardless of RMSD, reflecting the static MHG's insensitivity to conformational state. Under V2, ATE peaks at intermediate RMSD (approximately 11.6 Angstroms, frame 89) rather than at either endpoint or the maximum RMSD (20.8 Angstroms, frame 396), revealing that the harmonic network reaches its maximum collective coupling at a specific conformational state distinct from any geometric landmark.

**a**

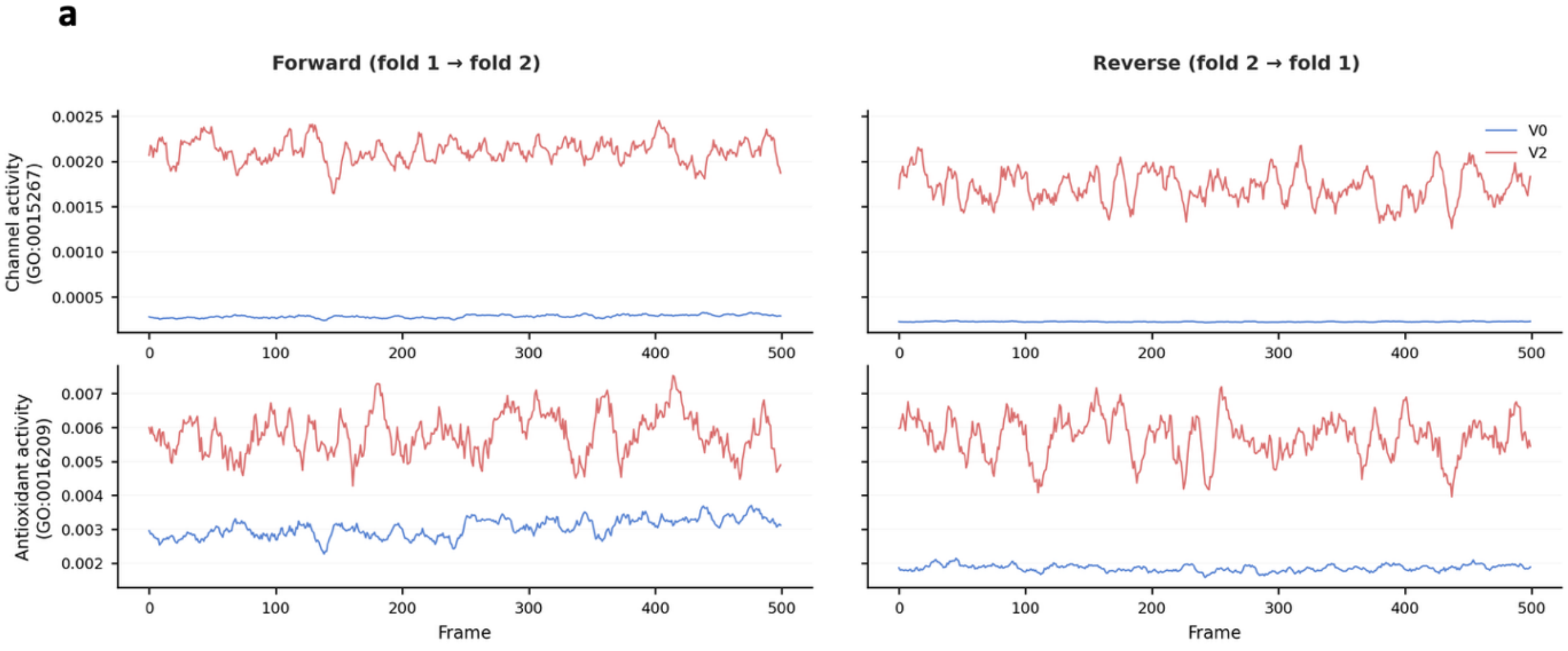

Forward (fold 1 → fold 2)
Reverse (fold 2 → fold 1)
Channel activity (GO:0015267)
Antioxidant activity (GO:0016209)
V0
V2
Frame

**b**

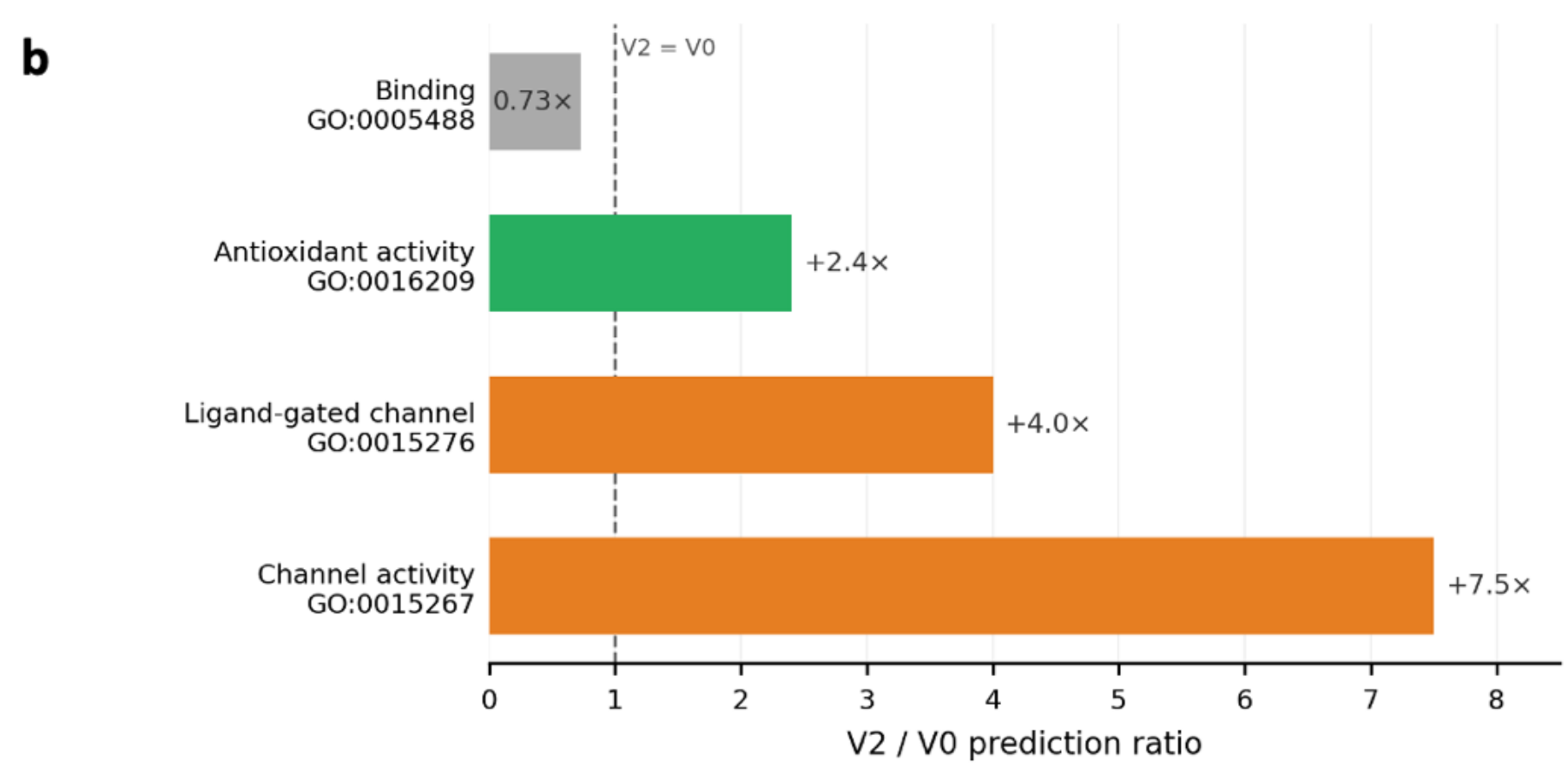

V2 = V0
Binding GO:0005488
0.73×
Antioxidant activity GO:0016209
+2.4×
Ligand-gated channel GO:0015276
+4.0×
Channel activity GO:0015267
+7.5×
V2 / V0 prediction ratio

**c**

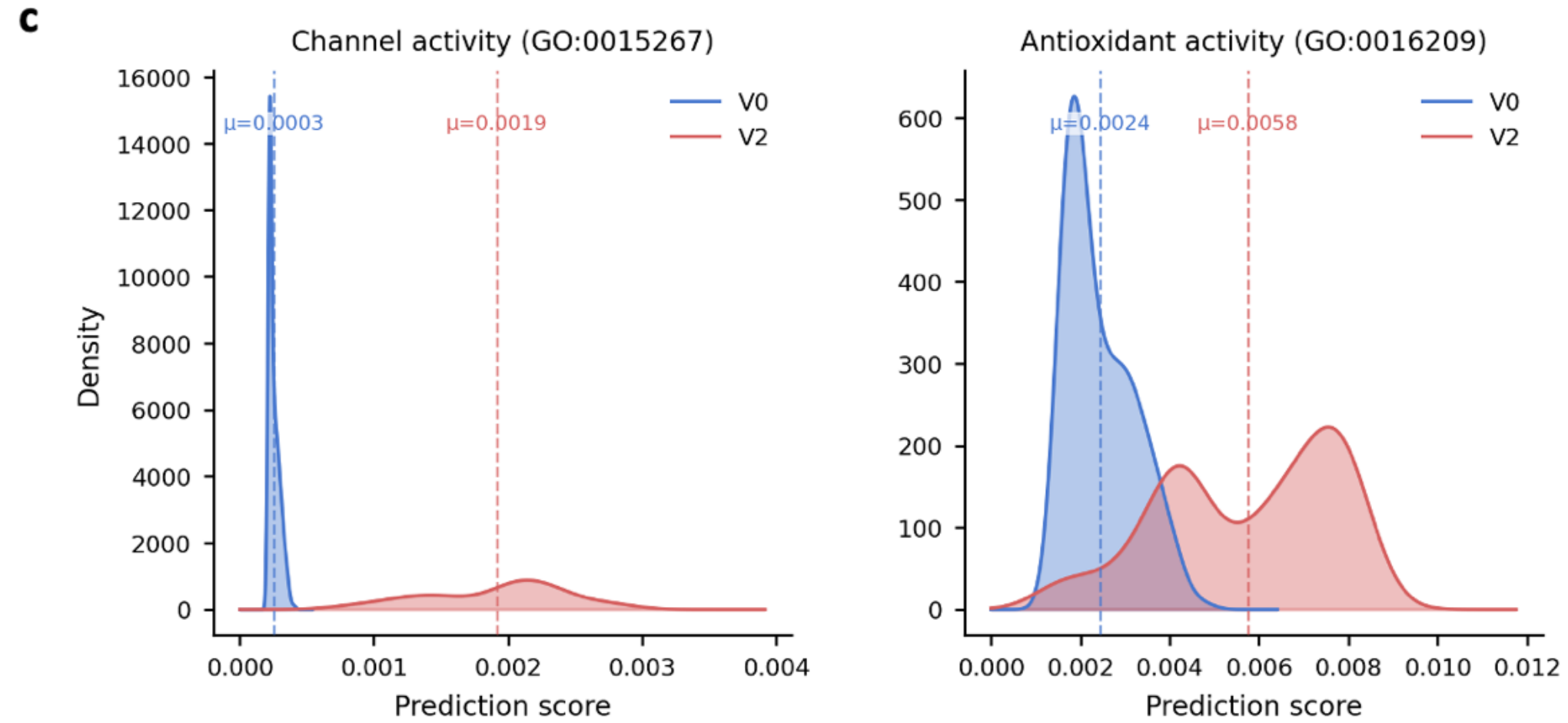

Channel activity (GO:0015267)
μ=0.0003
μ=0.0019
Antioxidant activity (GO:0016209)
μ=0.0024
μ=0.0058
V0
V2
Density
Prediction score

**Fig 5.**

**Figure 5. Kuramoto entrainment recovers both functional states of CLIC1 from vibrational dynamics alone.** (a) Per-frame GNN prediction scores for channel activity (GO:0015267) and antioxidant activity (GO:0016209) across the two 500-frame CLIC1 steered MD transition trajectories, for V0 static and V2 Kuramoto-entrained MHGs. V2 predictions for both functional states are systematically elevated relative to V0, with the effect most pronounced on channel activity where the fold 2 state is expected to produce the strongest signal. (b) Trajectory-averaged V2-over-V0 amplification ratios for four GO molecular function terms tracked across the trajectory. V2 amplifies channel activity prediction 7.5-fold, ligand-gated channel activity 4.0-fold, and antioxidant activity 2.4-fold, while the general binding term is unchanged or slightly reduced, confirming that amplification is specific to the functional categories associated with the protein's two folds rather than a uniform upward shift in prediction scores. (c) Distribution of per-frame channel activity and antioxidant activity predictions across the trajectory, V0 compared to V2. V2 produces wider, more bimodal distributions consistent with distinguishable predictions at the two fold endpoints, while V0 distributions are compressed near zero. CLIC1 was excluded from training by the 30% sequence-identity homology filter, and no sequence information was used at prediction time.

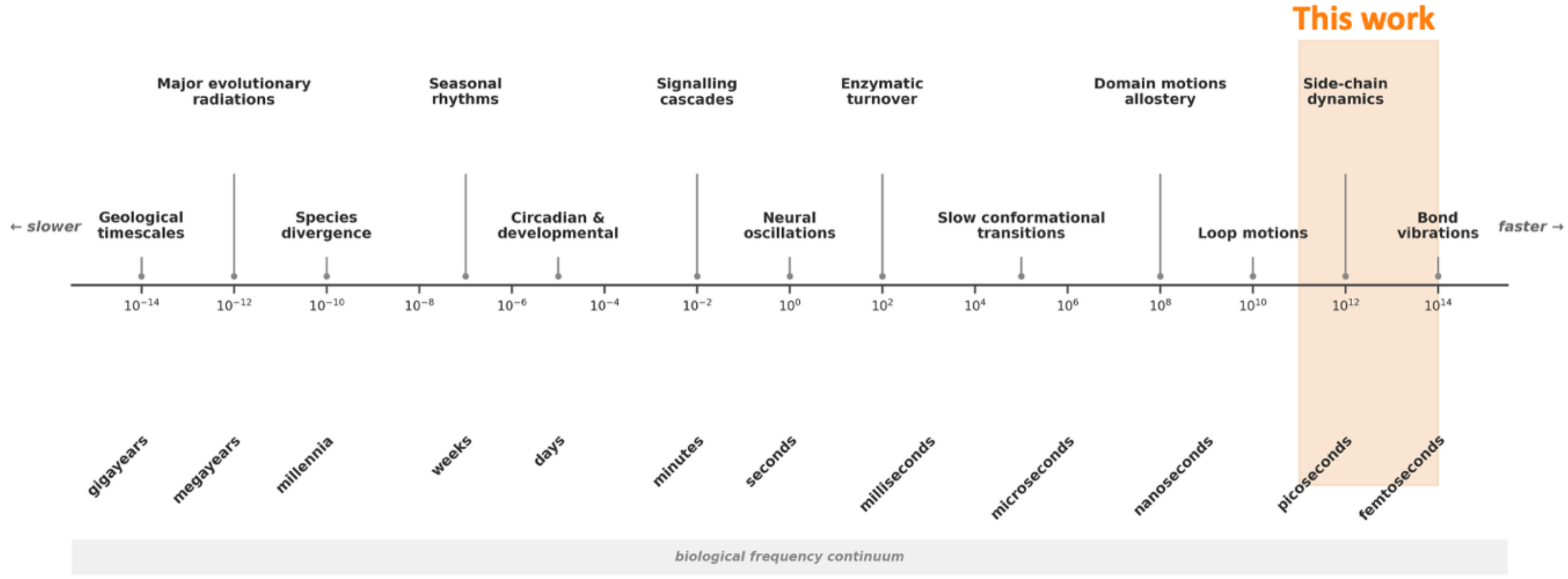

**Fig 6.**

**Figure 6. The biological frequency continuum, from bond vibrations to evolutionary timescales.** A logarithmic frequency axis spanning approximately 28 orders of magnitude maps characteristic biological processes to their dominant frequencies: femtosecond bond vibrations, picosecond side-chain dynamics, nanosecond loop motions, microsecond domain motions and allosteric transitions, millisecond conformational switching, second-scale enzymatic turnover, minute-to-hour signalling cascades, hour-to-day cell cycle dynamics, day-to-year developmental and physiological rhythms, and longer evolutionary timescales. The narrow band addressed in the present work (vertical highlight) corresponds to the 10-picosecond molecular dynamics trajectories used to construct the mechanical harmonics graphs of 5,238 proteins, approximately $10^{11}$ to $10^{14}$ Hz. All processes in the continuum admit covariance descriptions and therefore admit mechanical harmonics graph representations within the same formal structure. Frequency-space mechanics provides the formal infrastructure through which biological processes at all scales become comparable within a single representational space; the current paper is one entry point into this continuum.